# Nanostructured $Fe_2O_3/Cu_xO$ Heterojunction for Enhanced Solar Redox Flow Battery Performance

*Jiaming Ma[1], Milad Sabzehparvar[1], Ziyan Pan[1], Giulia Tagliabue[1]**

*[1] Laboratory of Nanoscience for Energy Technologies (LNET), STI, École Polytechnique Fédérale de Lausanne, 1015 Lausanne, Switzerland*

**Corresponding Author. Email: giulia.tagliabue@epfl.ch*

## Abstract

Solar redox flow batteries (SRFB) have received much attention as an alternative integrated technology for simultaneous conversion and storage of solar energy. Yet, the photocatalytic efficiency of semiconductor-based single photoelectrode, such as hematite, remains low due to the trade-off between fast electron hole recombination and insufficient light utilization, as well as inferior reaction kinetics at the solid/liquid interface. Herein, we present an $\alpha$-$Fe_2O_3/Cu_xO$ p-n junction, coupled with a readily scalable nanostructure, that increases the electrochemically active sites and improves charge separation. Thanks to light-assisted scanning electrochemical microscopy (Photo-SECM), we elucidate the morphology-dependent carrier transfer process involved in the photo-oxidation reaction at a $\alpha$-$Fe_2O_3$ photoanode. The optimized nanostructured is then exploited in the $\alpha$-$Fe_2O_3/Cu_xO$ p-n junction, achieving an outstanding unbiased photocurrent density of 0.46 $mA/cm^2$, solar-to-chemical (STC) efficiency over 0.35% and a stable photocharge-discharge cycling. The average solar-to-output energy efficiency (SOEE) for this unassisted $\alpha$-$Fe_2O_3$-based SRFB system reaches 0.18%, comparable to previously reported DSSC-assisted hematite SRFBs. The use of earth-abundant materials and the compatibility with scalable nanostructuring and heterojunction preparation techniques, offer promising opportunities for cost-effective device deployment in real-world applications.

# Introduction

Solar energy conversion offers a promising solution to meet the steadily increasing energy demand sustainably. Through the combination of photoelectrochemical cells (PEC) and redox flow batteries (RFB), solar energy can be efficiently converted and stored as chemical fuels by oxidizing or reducing various redox couples[1–3]. The success of this all-in-one solar redox flow battery (SRFB) mainly depends on the design of the cell structure[4,5] and the development of high-performance photoelectrodes[6,7].

Theoretically, the maximum solar-to-chemical (STC) efficiency of a single photoelectrode SRFB system can reach 16%-18% if the bandgap of the absorber material is within 1.4–2 eV and the thermodynamic cell voltage is around 0.9 V and 0.7 V[8]. Yet, practical realizations have not surpassed 3.9% STC efficiencies[9] with limited upscale fabrication. This is related to band-alignment constraints for SRFB photocharging up to high states of charge, which favor wide-bandgap materials with limited solar light absorption[10]. Additionally, low efficiency of charge separation dramatically reduces the performance of real devices. STC conversion efficiencies up to 21.1% have been instead achieved by exploiting complex device structures (e.g. dual photoelectrodes[11]) or more expensive tandem photoelectrodes (e.g. multi-junction solar cells[12,13]). Thus, the development of single photoelectrodes with low cost and scalable manufacturing necessitates further investigation[14].

Hematite ($\alpha$-$Fe_2O_3$) is a promising photoanode material due to its stability, non-toxicity, low cost, abundance on Earth, and attractive band gap[15,16] (1.9-2.2 eV). However, its performance is hindered by the trade-off between light-absorption and charge separation/transport, due to the short hole diffusion length[17] and poor charge carrier conductivity[6,18]. Hematite thin films[19,20] have been used to reduce the charge carrier diffusion distance to the electrode/electrolyte interface at the expense of complete light absorption (a thickness of 40-100 nm is needed to absorb 450-550 nm light[21]). An effective approach to overcome this trade-off is the use of nanoengineered structures, which can shorten the charge carrier transfer length[22–24], increase the electrochemically active surface area[25], achieve light trapping[26], and induce optical resonances within the active photocatalyst material itself[27]. Additionally, properly engineered heterojunction photoelectrodes, e.g. based on a p-n junction[28], can further improve the spatial separation of photogenerated electron-hole pairs[29], enhancing the photocatalytic activity. In the context of water-splitting or photosynthetic devices, a significant amount of effort has been devoted to proposing and designing heterojunctions aimed at efficiently extracting photo-holes from $\alpha$-$Fe_2O_3$ catalysts[15,30–33]. Yet, this approach has not been thoroughly explored in semiconductor-based SRFBs, despite major advantages in enabling higher state-of-charge and higher voltages during photocharging and discharging, respectively. Overall, nanoengineering and heterojunction design have a large untapped potential for improving single photoelectrode SRFB PEC performance.

In this work, we present a scalable, nanostructured α-$Fe_2O_3$/$Cu_xO$ p-n junction and demonstrate its largely improved unassisted photocharging of an integrated solar redox flow battery (**Figure 1a**). First, α-$Fe_2O_3$/$Cu_xO$ films with varying thicknesses were systematically investigated to elucidate the impact of the p-n junction on the photoelectrochemical performance (**Figure 1b**). Concurrently, light-assisted scanning electrochemical microscopy (Photo-SECM) was employed to reveal enhanced charge separation in α-$Fe_2O_3$ nanopillar arrays (**Figure 1c**). Finally, guided by the results from SECM, the optimized α-$Fe_2O_3$ nanostructure was integrated with the p-n junction strategy to enhance charge carrier separation while improving electrochemically active sites, thus resulting in a high performance photoanode. Our SRFB, featuring the nanostructured α-$Fe_2O_3$/$Cu_xO$ p-n junction, demonstrates record values of unassisted photocurrent (0.46 mA/ $cm^2$), along with STC efficiency ~0.35% and SOEE ~0.18%, comparable to solar cell-assisted hematite-based devices. Overall, this α-$Fe_2O_3$-based SRFB shows a stable photocharge-discharge cycling performance and presents opportunities to drive real-world deployment of more cost-effective devices.

# Experimental Section

## Sample Preparation

**Synthesis of α-$Fe_2O_3$/$Cu_xO$ Film photoanodes** (**Figure S1a**, Supporting Information)**.** RF magnetron sputtering was used to sputter iron thin films (15, 30 and 50 nm) on Indium Tin Oxide (ITO) glass under the protection of Argon gas. Subsequently, copper thin films with different thickness (15, 30 and 45 nm) were sputtered on the iron films under the same condition. The Fe/Cu films were then immersed in 4 M NaOH (1 h at 80 °C, then 20 h at room temperature) to get Fe/$Cu_2O$ film in accordance with previous reports[34]. Next, all the samples were annealed at 645 °C under air for 10 minutes with the ramping of 5 °C/min to obtain α-$Fe_2O_3$/ $Cu_xO$ (named F-$Cu_xO$) photoanodes. The control samples, named 15F, 30F, and 50F, consisted of iron films with thicknesses of 15 nm, 30 nm, and 50 nm, respectively, without copper coating or NaOH treatment, all subjected to the same annealing process, which turns them into hematite films.

**Synthesis of α-$Fe_2O_3$/CuO Film photoanodes.** Iron thin film (15 nm) was sputtered on Indium Tin Oxide (ITO) glass by using the same method mentioned above. Subsequently, copper thin (30 nm) was sputtered on the iron film under the same condition. Then, the sample was annealed at 645 °C under air for 10 minutes with the ramping of 5 °C/min to obtain α-$Fe_2O_3$/ CuO (named 15/30 F-CuO) photoanodes.

**Synthesis of α-$Fe_2O_3$ nanopillar array** (**Figure S1b**, Supporting Information)**.** Iron thin film (30 nm) was sputtered on Indium Tin Oxide (ITO) glass by using the same method mentioned above. The Fe nanopillars were then made by e-beam lithography and ion beam etching. ZEP-520A (50%) was spin coated at 4000 rpm rate (~120 nm) on a precleaned iron sample, followed by baking at 180°C for 5 minutes. E-beam was used to pattern nanostructures on photoresist and the sample was subsequently developed in Amyl-Acetate solution for 1 minute. The desired Fe nanopillars (300 nm periodicity; 100, 150 and 200 nm in diameter) was then fabricated by ion beam etching with 1.1 nm/s etching speed. The as-prepared Fe nanostructure consists of nanopillars that are 25 nm in height, with a continuous Fe layer that is 5 nm thick at the bottom. The sample was cleaned via oxygen plasma (150 sccm $O_2$, 200 W) for 10s, followed by annealing at 645 °C under air for 10 minutes to obtain α-$Fe_2O_3$ nanopillar array (named P100, P150 and P200).

**Synthesis of nanostructured α-$Fe_2O_3$/$Cu_xO$ photoanode** (**Figure S1c**, Supporting Information)**.** Iron thin films (30 nm) were sputtered on Indium Tin Oxide (ITO) glass as mentioned above. Polystyrene (PS) nanospheres (Microparticles GmbH) with average diameter of 300 nm were then coated on top of iron film via Langmuir Blodgett (LB) technique as a monolayer[35]. Oxygen plasma (800 sccm $O_2$, 300 W) was used to reduce the PS nanospheres to a diameter ranging from 150 to 180 nm. The PS etching speed is around 7.5 nm/min. Subsequently, similar Fe nanopillars with Fe thin film (~5 nm) was obtained via ion beam etching and beads removing process[36]. 30 nm copper film was then sputtered on the Fe nanostructure. The same NaOH treatment and annealing process as mentioned above were performed to get the nanostructured α-$Fe_2O_3$/$Cu_xO$ (named P/$Cu_xO$) photoanode. The bare nanostructured α-$Fe_2O_3$ (named P) without copper coating and NaOH treatment was also fabricated as a control sample via the same method.

The summary of the sample preparation process is shown in Table S1.

## Material Characterization

The morphology and crystal structure of photoanodes were characterized by a scanning electron microscope (Zeiss Gemini SEM 300) and X-ray diffractometer (XRD, Rigaku Synergy-I single crystal). Ultraviolet photoelectron spectroscopy (UPS) was performed using a PHI VersaProbe II scanning XPS microprobe (Physical Instruments AG, Germany) equipped used with He(I) and He(II) UV source.

## Optical Measurements

The UV-Vis test for α-$Fe_2O_3$/$Cu_xO$ Film and nanostructured α-$Fe_2O_3$/$Cu_xO$ were performed under a solar simulator (Newport 66984-300XF-R1 Xe lamp) with an AM 1.5G filter as the light source, using a monochromator (Newport, CS260B-2-MC-A) connected with an integrating sphere (Newport, 819D-IS-5.3). The absorption data was obtained following our previous work[37].

For the α-$Fe_2O_3$ nanopillars, an inverted microscope (Nikon Eclipse Ti2) was used in combination with a grating spectrometer (Princeton Instruments Spectra Pro HRS-500) equipped with a Peltier-cooled 2D CCD detector (Princeton Instruments PIXIS 256) to record reflection (R) and transmission (T) spectra. The absorption (A) was determined as A=1-R-T. The detailed process is in accordance with previous reports[38,39].

## Photoelectrochemical Measurements and Photocharge-Discharge

Linear Sweep Voltammetry (LSV), Photocurrent density-time (j-t) and Electrochemical Impedance Spectroscopy (EIS) were used to evaluate the photoelectrochemical performance of different photoanodes and were recorded on a Biologic SP-300 potentiostat. LSV were tested both in dark and under illumination with a scan rate of 10 mV/s; J-t was recorded without bias and the photo response signal were obtained with 20-20 s light on/off. EIS analyses were carried out at a perturbation amplitude of 10 mV with the frequency ranging from 0.01 Hz to 10 kHz. EC-Lab software was used to fit the measured EIS results. All tests were carried out by using a two-electrode configuration of the SRFB device (one photoanode in anolyte and one 39 AA carbon felt in catholyte) under 1 sun illumination (AM 1.5 G filter, 100 mW/cm$^2$). All samples were back-illuminated through the ITO glass, and the illuminated area is 0.785 cm$^2$.

The photocharge-discharge behavior of our solar redox flow batteries was demonstrated via three electrodes integrated SRFB and was recorded by two potentiostats. During photocharging process, potentiostat 1 (SP-300) was connected to the PEC part to monitor the photocurrent, while potentiostat 2 (CHI 760E), connecting the RFB part, measured the evolution of the cell potential. During the discharging process, the solar simulator and potentiostat 1 were turned off, while a discharging current of 0.4 mA was applied by potentiostat 2 to RFB part until the cell potential reached 0 V.

During all these tests, the electrolyte recirculation was guaranteed by using two peristaltic pumps (30 ml/min), both the anolyte (0.2 M $Na_4Fe(CN)_6$/1 M NaOH) and catholyte ( 0.1 M 2, 7-AQDS/1 M NaOH ) volume of the running system were 4 ml.

## Scanning Electrochemical Microscopy

Light-assisted scanning electrochemical microscopy (Photo-SECM) was implemented to investigate the photocatalytic activity of micro-array structures (100 x 100 µm$^2$) of α-$Fe_2O_3$ nanopillars, using a previously described home-built instrument[38]. Briefly, a home-built SECM was coupled with an inverted optical microscope (Nikon Eclipse Ti2) and a white light source (Energetiq EQ-99X-FC LDLS) for back-illumination of the structures. A three-electrode configuration was employed with Pt and Ag/AgCl as counter and reference electrodes, and Pt ultra-microelectrode (Pt UME) tip as the working electrode. The structured samples were unbiased and grounded. Tip to substrate distance was controlled by monitoring the tip current during its fine approach towards the substrate and lifting up the tip by 3 um higher than the distance corresponding to 30% change from the bulk state. All experiments were performed in a 4mM $K_4Fe(CN)_6^{4-}$ and 0.4M KOH solution, using a 1.2 µm radius UME tip (RG value=14.5) biased at a reductive 0V vs Ag/AgCl tip potential, and modulating a ~ 90 µm diameter collimated light beam having 80 mW/m$^2$ power density.

## Numerical simulation

Electromagnetic simulations were performed using the RF module of COMSOL Multiphysics v6.1 to obtain the absorption spectra of α-$Fe_2O_3$ nanopillars and α-$Fe_2O_3$ film. For α-$Fe_2O_3$ nanopillars, a 3D unit cell model with periodicity of 300 nm, consisting of one α-$Fe_2O_3$ nanopillar (50 nm in height) on top of 10 nm α-$Fe_2O_3$ film/100 nm ITO film /fused silica substrate surrounded with a top layer of air, was simulated by setting the diameter of nanopillars from 100 nm to 250 nm with 10 nm step size. For α-$Fe_2O_3$ film simulations, a similar 3D unit cell model without the α-$Fe_2O_3$ nanopillar was performed by varying the film thickness from 20 – 180 nm with 10 nm step size. In both cases, perfect magnetic conductor and perfect electric conductor boundary conditions were used at the side walls of the unit cell. A port boundary condition was used at the bottom of the unit cell. The back illumination was applied with a normal incident plane wave (300-850 nm) with electric field polarization perpendicular to the film plane as well as recording the reflected wave. At the top of the unit cell, a second port boundary condition without excitation was used to record the transmitted wave. The refractive indices for α-$Fe_2O_3$ and ITO were taken from literatures[40,41]. Then the absorbed power was calculated by volume integration of the electromagnetic power loss density over the α-$Fe_2O_3$ volume.

# Results and Discussion

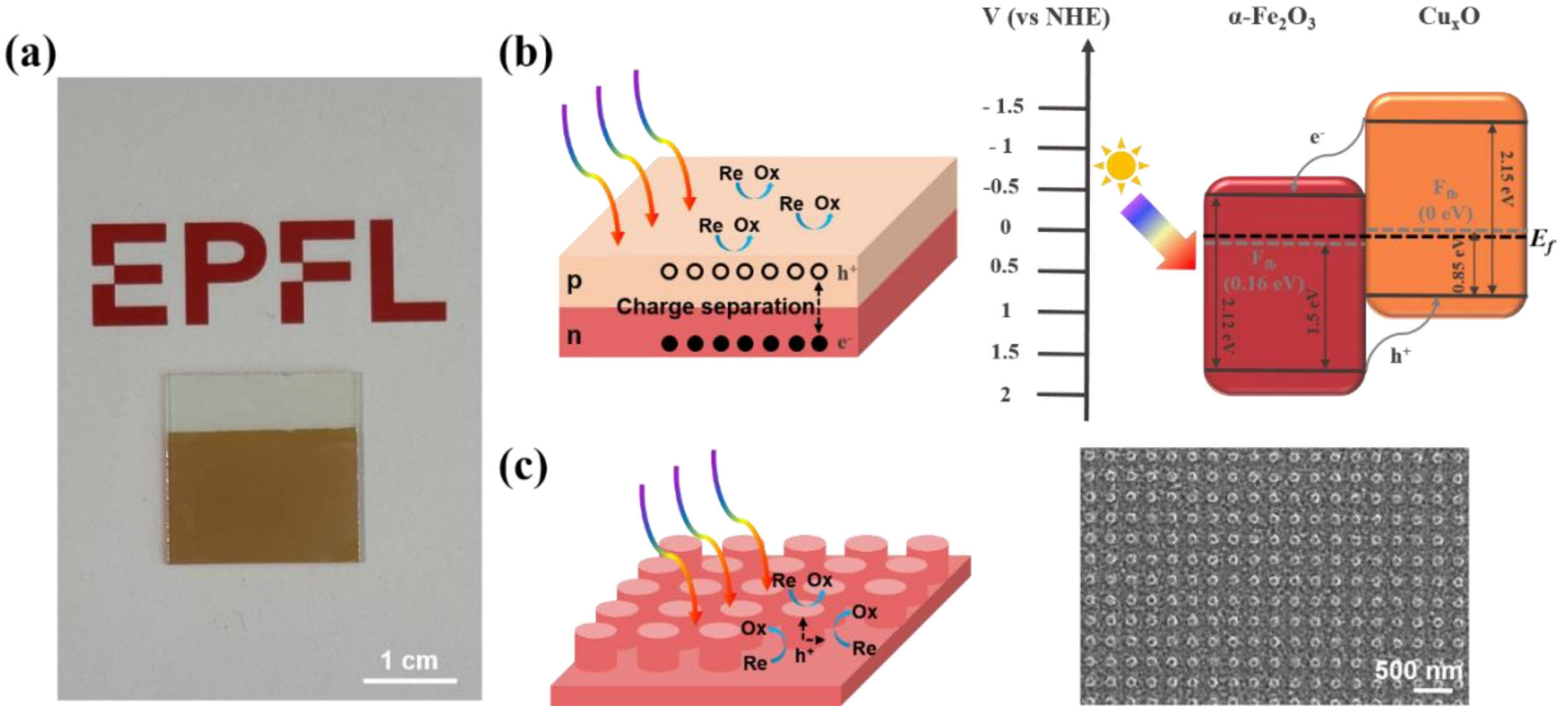

**Figure 1.** (a) Digital figure of a  P/CuₓO sample (5 cm$^2$), obtained by synergistically combining heterojunction engineering with large-area nanofabrication (nanosphere lithography; (b) Schematic of a planar p-n junction and its contribution to electron-hole separation (left) as well as quantitative band alignment of the as prepared α-$Fe_2O_3$ and $Cu_xO$ (right); (c) Schematic a nanostructured photoanode and its improved hole collection (left) and SEM of one of the studied α-$Fe_2O_3$ nanopillars (right, period of 300 nm and diameter of 150 nm).

α-$Fe_2O_3$/$Cu_xO$ heterojunctions were fabricated via a facile and scalable method to enhance electron/hole separation in SRFB photoanodes. Specifically, after sputtering 15 nm Fe and between 15 nm and 45 nm Cu onto ITO, a NaOH treatment was used to convert Cu to $Cu_2O$[34]. Subsequently, annealing of the as-prepared composite leads to the formation of α-$Fe_2O_3$-$Cu_2O$-CuO, concisely referred to α-$Fe_2O_3$/$Cu_xO$ (see Experimental Section). In the following, samples are named according to the thickness of the initial Fe and Cu film (e.g. 15/30 F-$Cu_xO$ for the heterostructure). We note that an approximately 2-fold expansion is expected for conversion from Fe to $Fe_2O_3$[42] while a significant reduction in thickness is expected for the Cu to $Cu_xO$ conversion duo to the partially dissolving of Cu into NaOH[34]. As a control sample, we use a 15 nm Fe film that is annealed under the same conditions without NaOH treatment (sample 15F, see Experimental Section).

**Figure 1b** shows the estimated band alignment for our α-$Fe_2O_3$/$Cu_xO$ heterojunction, obtained by combining the measured bandgap from Tauc plot, work function from the Mott-Schottky technique and the valence band maxima from UPS (**Figure S2**, Supporting Information). This is indeed crucial to assess the energy levels compatibility between the electrode/electrolyte. We confirmed that $Cu_xO$ is a p-type semiconductor and that a p-n junction is formed at the α-$Fe_2O_3$/$Cu_xO$ interface. While this is expected to promote charge separation, and hence the photoelectrochemical performance of the photoanode[29], it can reduce the possible theoretical discharge cell voltage (**Figure 1b**), as the oxidation and reduction potentials of the chosen redox couples must lie within the p-type semiconductor valence band and the n-type semiconductor conduction band energies[29]. We chose $Fe(CN)_6^{4-}/Fe(CN)_6^{3-}$ as the anolyte (0.60 V vs normal hydrogen electrode (NHE)) and $AQDS/AQDS^{2-}$ (-0.14 V vs NHE) as the catholyte to evaluate the performance of the designed α-$Fe_2O_3$-based photoanodes[5].

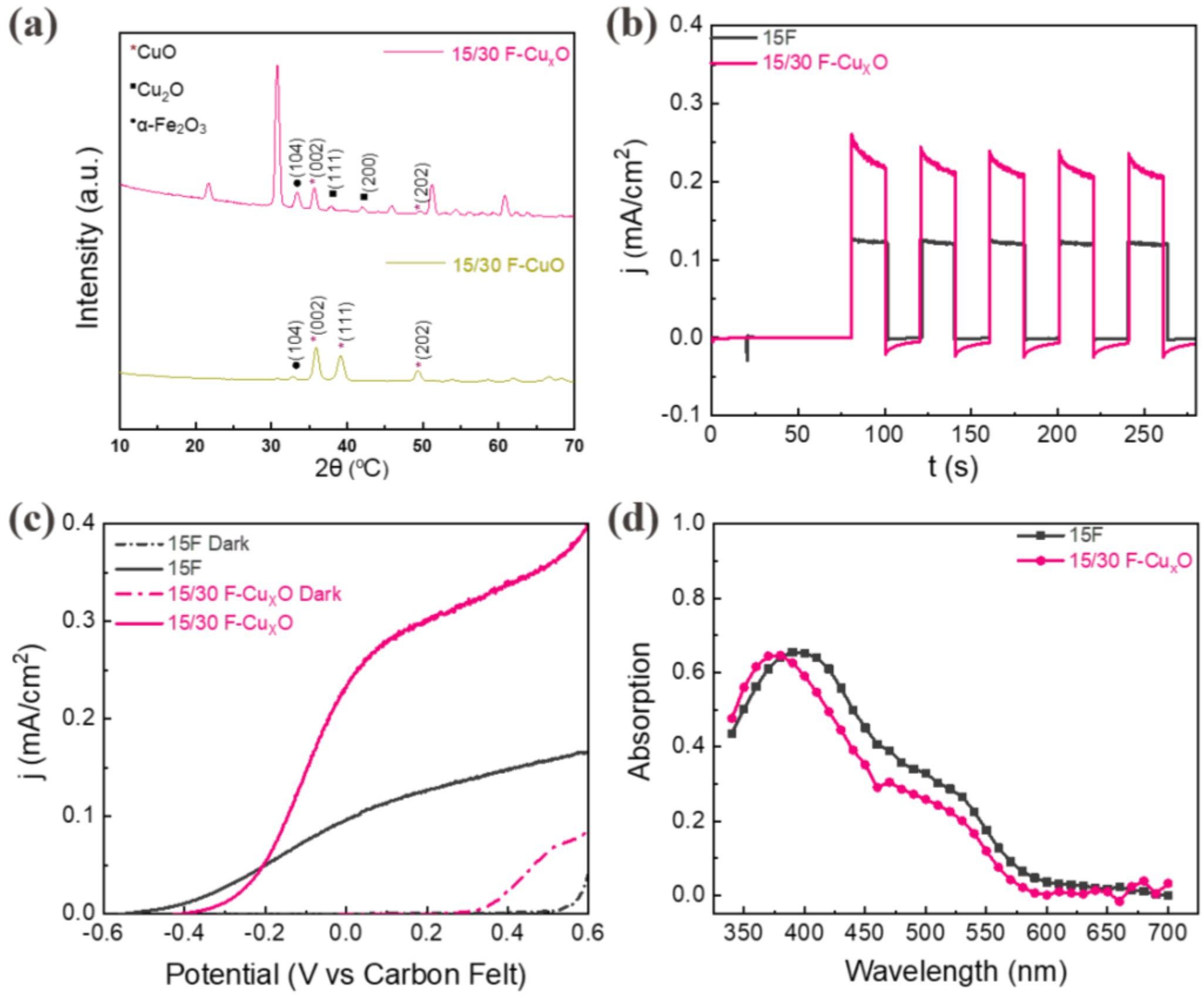

**Figure 2.** (a) XRD of different photoelectrodes; (b) Photoresponse behavior for 15/30 F-$Cu_xO$ and 15F with 20 s light on-off; (c) Linear sweep voltammetry curve for 15/30 F-$Cu_xO$ and 15F both under 1 sun illumination and dark; (d) UV-Vis absorption spectrum of 15/30 F-$Cu_xO$ and 15F.

A two-electrode SRFB was used to perform photocurrent density-time (j-t) measurements and investigate the PEC performance of the photoanodes under 1 sun illumination. The NaOH treatment time and the thickness of sputtered Cu have been optimized, identifying 15/30 F-$Cu_xO$ as the best p-n junction film (**Figure S3**, Supporting Information). XRD measurements (**Figure 2a**, pink curve) confirm that the 15/30 F-$Cu_xO$ sample present the hematite phase of α-$Fe_2O_3$ and a mixture of $Cu_2O$-CuO ($Cu_xO$). The sharp diffraction peaks at 2θ of 33.1° indicate the (104) plane of the rhombohedral structure of hematite[43]. The peaks observed at 35.4° and 48.8° correspond to the (002) and (202) planes of CuO, while the peaks at 36.6° (111) and 42.2° (200) are characteristic peaks of $Cu_2O$[44]. The rest of the peaks can be attributed to ITO[45]. Additionally, we observed that excluding the NaOH treatment of the Cu film (sample 15/30 F-CuO, see Experimental Section) results only in CuO diffraction peaks without any trace of $Cu_2O$ (**Figure 2a**, green curve). From a morphological point of view, scanning electron micrographs (**Figure S4**, Supporting Information) interestingly show that the 15/30 F-$Cu_xO$ heterojunction film exhibits a smoother surface and more uniform grains than the control hematite sample 15F. This can be attributed to the formation of $Cu_xO$ during the annealing process of $Cu_2O$ on top of α-$Fe_2O_3$.

As shown in **Figure 2b**, the unbiased photocurrent density for 15/30 F-$Cu_xO$ is 0.24 mA/cm$^2$, two times higher than the control hematite sample (0.12 mA/cm$^2$). Both 15/30 F-$Cu_xO$ and 15F show a stable pulse signal with instantaneous photoresponse, indicating their excellent photoactivity. The long-term operation (1 h) of 15/30 F-$Cu_xO$ also exhibits remarkable photocurrent stability, with a current retention of approximately 98% (**Figure S5**, Supporting Information), indicating its robust structure as well as enduring photocatalytic stability. Additionally, linear sweep voltammetry (LSV) curves of these two photoanodes were measured in the same set up both under illumination and dark conditions with a sweeping rate of 10 mV/s (**Figure 2c**). Indeed, the photocurrent onset of 15/30 F-$Cu_xO$ is observed around -0.42 V vs carbon felt, slightly higher than that of 15F (-0.52 V), owing to the partial sacrifice of oxidation and reduction potentials by the p-n junction as discussed above. Given their consistent dark current onset (0.3 V), the photovoltage (defined here as the potential difference between dark and light current onset) of 15/30 F-$Cu_xO$ is smaller than that of 15F as expected. The unbiased photocurrent at 0 V in linear sweep voltammetry of these two photoanodes are in line with j-t tests, further demonstrating the significantly increased photocatalytic activity of the heterojunction.

When considering solely the photoelectrode components (i.e. excluding considerations of battery resistance losses and redox couple reaction activities), the theoretically achievable photocurrent is decided by the light absorption and the charge carriers transfer from the bulk to the electrode/electrolyte interface of semiconductors[37]. The optical response of the bare α-$Fe_2O_3$ film was measured and compared with that of the 15/30 F-$Cu_xO$ in **Figure 2d**. The UV absorption (340-400 nm) increases from 15F to 15/30 F-$Cu_xO$, while the visible light absorption (400-600 nm) of 15F is slightly higher than 15/30 F-$Cu_xO$. Overall, with their total light absorption spectra being comparable, the primary factor influencing photocurrent becomes the photogenerated electron-hole separation process, indicating that the 15/30 F-$Cu_xO$ p-n junction exhibits better charge transfer than 15F. As a comparison, we tested thicker hematite films (30F and 50F) with higher absorption as well as their p-n junction counterparts (**Figure S6-S8**, Supporting Information). Yet, in all cases due to the limited charge mobility within hematite[17], the measured photocurrent was lower than for the 15/30 F-$Cu_xO$ film. Electrochemical Impedance Spectroscopy (EIS) analysis (**Figure S9**, Supporting Information) further confirmed the photocurrent measurements. The internal charge transfer resistance ($R_{sc}$) of 15/30 F-$Cu_xO$ is 1530 Ω, significantly lower than that of 15F (5430 Ω), suggesting efficient charge carrier transport within the bulk facilitated by the p-n junction. Additionally, the charge transfer resistance at the interface ($R_{ct}$) decreases from 15F (4060 Ω) to 15/30 F-$Cu_xO$ (1050 Ω), indicating faster photooxidation reaction dynamics at the 15/30 F-$Cu_xO$/ferrocyanide interface compared to the 15F/ferrocyanide interface. Hence, these two factors contribute to the higher photocurrent density of the planar 15/30 F-$Cu_xO$ p-n junction.

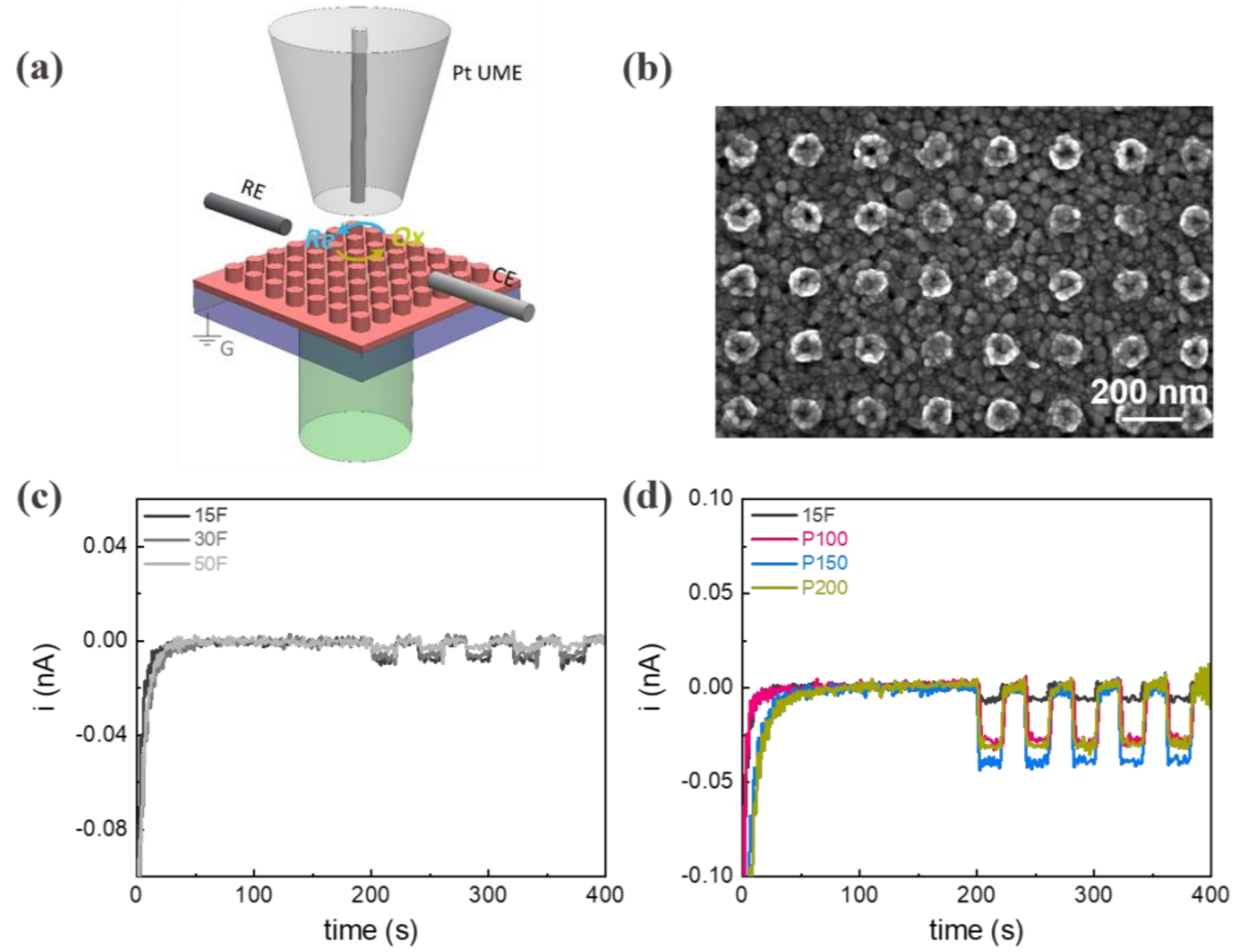

**Figure 3.** *(a)* Schematic of light-assisted scanning electrochemical microscopy (Photo-SECM) set-up; (b) SEM of P150; SECM to probe α-$Fe_2O_3$ (c) films (F) and (d) nanopillar arrays (P) under white light illumination (80 mW/$cm^2$).

To improve charge transfer in thicker hematite films that exhibit higher light absorption (**Figure S6,** Supporting Information), we explored the impact of nanoengineering strategies, which increase the available surface area and reduce the charge transport distances. Specifically, we realized arrays of Fe nanopillars with well-defined diameter, *D*, and periodicity, *P*, which were subsequently oxidized to obtain α-$Fe_2O_3$ nanopillars. A thin (~10 nm) α-$Fe_2O_3$ film was left to cover the ITO substrate (see **Figure S1b** and Experimental Section). Electromagnetic simulations (COMSOL Multiphysics®) were used to quantify the impact of *D* on the absorption spectrum of α-$Fe_2O_3$ nanopillar arrays based on a 30 nm thick iron (*P* = 300 nm). We observe that, despite the reduction in overall absorbing material, nanostructuring allows the excitation of optical resonance modes that entail a high absorption level (**Figure S7b**, Supporting Information). In order to explore the optimal balance between optical performance and electron transfer, α-$Fe_2O_3$ nanopillar arrays with diameters of 100, 150, and 200 nm (named as P100, P150, and P200 respectively) were patterned via E-beam lithography over an area of approximately 100 x 100 $um^2$. The SEM top-view images of nanopillar arrays with different diameters are reported in **Figure 3b** and **Figure S10a-f** (Supporting Information), showing the increase in the pillar diameter upon annealing, due to oxygen incorporation. Microscale absorption measurements[39] of the different arrays (**Figure S11**, Supporting Information) were consistent with simulations and showed a minimal decrease in absorption compared to the unpatterned film.

To investigate the effect of nanopatterning on the photocatalytic performance of the α-$Fe_2O_3$ photoanodes, we performed Photo-SECM on both films and nanopillar array structures in a 4 mM $Fe(CN)_6^{4-}$ and 0.4 M NaOH electrolyte solution under white light illumination. The reductively biased Pt UME tip in our experiments locally detects the photo-generated oxidant species, i.e. $Fe(CN)_6^{3-}$ (**Figure 3a**). For all the measurements, we controlled the tip to substrate distance by monitoring tip current versus distance and positioning the tip at a distance 3 um higher than the 30% offset value[46] (**Figure S1**2, Supporting Information). **Figure 3c** shows the time trace of the tip current for the α-$Fe_2O_3$ films having different thicknesses when the light is modulated on and off at the same power density. Analyzing the $I_{T,ON}/I_{T,OFF}$ values shows that the photocatalytic activity of the film structures decreases by increasing the film thickness from 15F to 30F, and then to 50F. This is in agreement with the lower charge transfer rates observed in the SRFB photoresponse as film thickness increases (**Figure S8a**, Supporting Information). Comparing the $I_{T,ON}/I_{T,OFF}$ values for the α-$Fe_2O_3$ nanopillar arrays (**Figure 3d**) with the best performing film structure (15F) shows a significant enhancement (more than 5 folds) in photocatalytic activity, not achievable by only decreasing the film thickness. Most importantly, the results show that the P150 α-$Fe_2O_3$ nanopillar array has the highest photocatalytic activity among all the samples, realizing an optimum combination of surface area, light absorption, and charge transport dynamics[27]. Based on these local PEC results, we chose 150 nm as the optimum Fe nanopillar diameter for realizing centimeter-scale, nanostructured α-$Fe_2O_3/Cu_xO$ photoanodes for SRFBs.

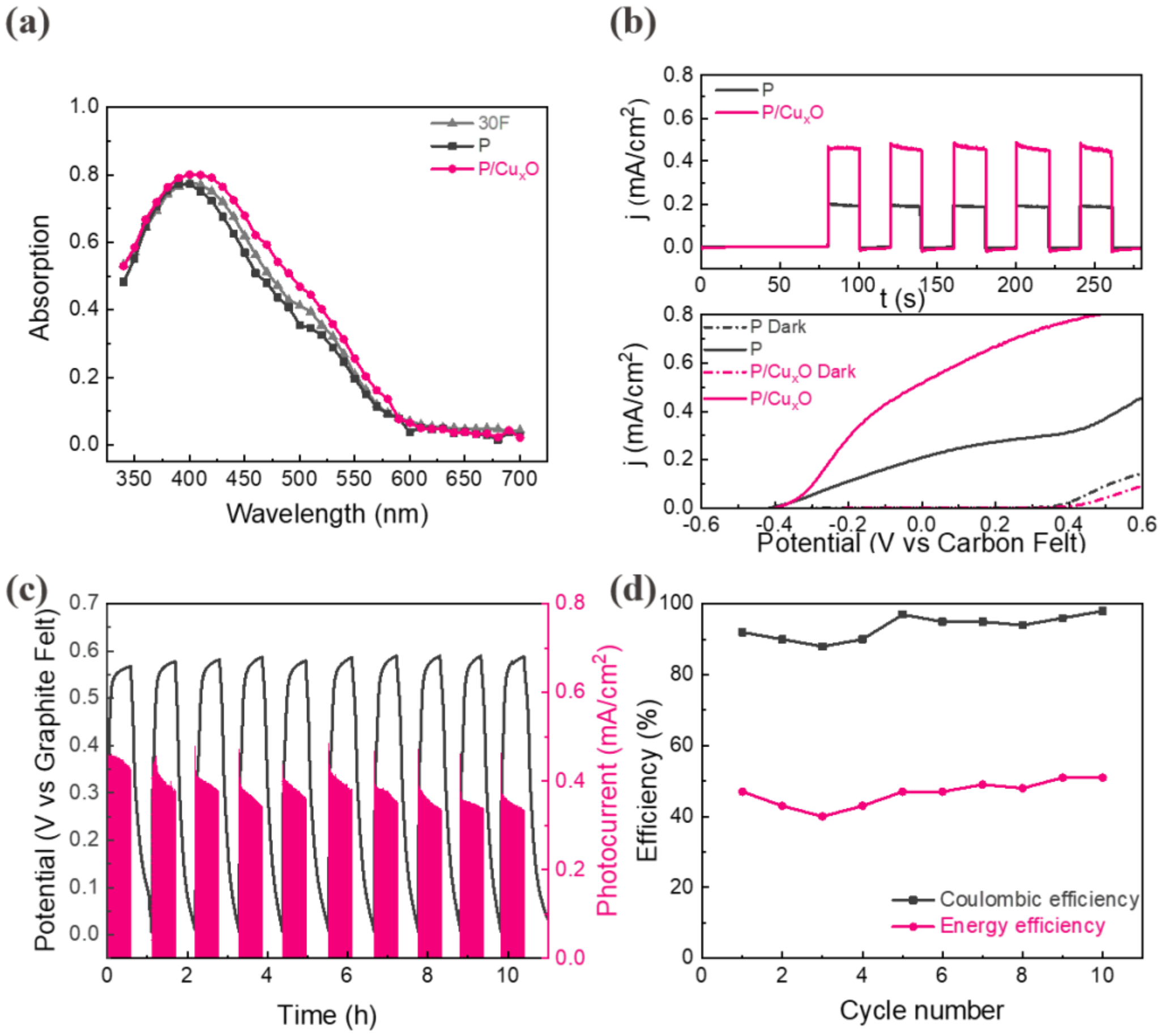

**Figure 4.** (a) UV-Vis absorption spectrum of nanostructures and 30F; (b) Photoresponse behavior for P/$Cu_xO$ and P with 20 s light on-off; Linear sweep voltammetry curve for P/$Cu_xO$ and P both under 1 sun illumination and dark; (c) Representative photocharge-discharge cycling behavior with a cut-off potential set to 0-0.58 V and unbiased photocurrent using P/$Cu_xO$ photoanode; (d) Coulombic and energy efficiency of photocharge/ discharge curve.

The nanostructured α-$Fe_2O_3$/$Cu_xO$ (P/$Cu_xO$) photoanode was fabricated employing a combination of nanoengineering and heterojunction design, as illustrated in **Figure S1c** (Supporting Information). A Langmuir Blodgett (LB) technique, rather than E-beam lithography, was exploited to fabricate nanopatterns over cm-scale samples[47]. A nanostructured α-$Fe_2O_3$ (P) without copper coating nor NaOH treatment was also fabricated as a control sample. The addition of the $Cu_xO$ layer results in distinct morphological changes. Specifically, the P/$Cu_xO$ exhibits nanorod bundles-like structures on the surface which are absent on the P sample (**Figure S13**, Supporting Information). Optically, the P sample exhibits the same light absorption of the unpatterned film while the P/$Cu_xO$ sample exhibits a small but broadband absorption increase (**Figure 4a**).

Photoelectrochemically, the nanostructured α-$Fe_2O_3$ (P) shows a photocurrent density of 0.2 mA/cm$^2$ (**Figure 4b**), nearly two folds higher than the best thin film sample, 15F, consistent with the SECM analysis.

With the addition of the p-n junction, which improves both light harnessing and carrier transport, the $P/Cu_xO$ sample exhibits an outstanding photocurrent density of 0.46 $mA/cm^2$ as well as a good stability (**Figure S14**, Supporting Information). **Figure 4b** illustrates the variation of photocurrent density with applied voltage. At 0V, the photocurrent density for $P/Cu_xO$ and P are 0.49 $mA/cm^2$ and 0.21 $mA/cm^2$, respectively. The higher photocurrents compared to the photoresponse tests may arise from the transient current generated by the double-layer capacitance under voltage changes as well as the trapped photogenerated holes due to the existence of detrimental surface states at the electrode–electrolyte interface[48]. Interestingly, the similar photovoltage (0.75 V) of both photoanodes suggest that the nanostructured p-n junction does not compromise its capability to drive these redox couples. EIS studies were also carried out under illumination to gain insight into the effect of the p-n junction on PEC redox oxidation reaction. The same electrical circuit was used as a model to fit the photooxidation process in the photoanodes (**Figure S15**, Supporting Information). In contrast to P, the $P/Cu_xO$ exhibits significantly lower values of $R_{sc}$ (500 Ω) and $R_{ct}$ (757 Ω), indicating that the surface coverage of an additional $Cu_xO$ layer can improve the charge separation inside bulk $\alpha$-$Fe_2O_3$ as well as reduce the charge extraction barrier to create a facile carrier pathway at the electrode/electrolyte interface. Additionally, the space-charge capacitance ($C_{sc}$) decreases from $P/Cu_xO$ ($1.4\times10^{-3}$ F) to P ($0.03\times10^{-3}$ F), implying the broadening of the depletion layer of the P photoanode and thus much inferior carrier mobility[49–51].

After demonstrating the photooxidation reaction activity and stability of the $P/Cu_xO$ photoanodes, the photocharge/discharge test was finally performed using a fully integrated SRFB. The photocharge/ discharge curves of the SRFB for the initial 10 cycles (approximately 11h) are shown in **Figure 4c**. The average photocurrent density is around 0.42 $mA/cm^2$ (0.33 mA when considering the active area) under 1 sun illumination, with a discharging current applied of -0.4 mA. The almost identical cycles imply the stability of both photoanodes and the overall SRFB system. This integrated system exhibits a high coulombic efficiency around 90%-98% and an average energy efficiency around 50% as shown in **Figure 4d**. Overall, the $P/Cu_xO$-based SRFB achieves a stable solar-to-chemical efficiency of 0.35% and an average solar-to-output energy efficiency of 0.18% over 12 cycles, which is a significant progress for unassisted $\alpha$-$Fe_2O_3$-based SRFB.

# Conclusion

In summary, the synergistic design of nanostructuring and engineered heterojunction for photoelectrodes was realized to improve the SRFB performance. Firstly, we conducted a comprehensive investigation through band alignment engineering and various electrochemical techniques, elucidating the enhancement of carrier transport both within the bulk photoelectrode and at the photoelectrode/electrolyte interface facilitated by the $Cu_xO$-$Fe_2O_3$ p-n junction rather than CuO-$Fe_2O_3$. Secondly, leveraging the localized measurements of SECM, we are able to utilize microscale samples to seamlessly explore the nanostructure sizes effect on the photoelectrodes performance. Nanoengineering not only enables the manipulation of the optical properties of the photoelectrodes (including light trapping and Mie resonance), but also provides additional reactive sites, effectively mitigating losses attributable to charge recombination. Indeed, by combining these two strategies, the nanorod bundles-like P/$Cu_xO$ exhibits the highest unbiased photocurrent density (0.46 mA/cm$^2$) as well as good stability for unassisted α-$Fe_2O_3$-based SRFB. The average STC efficiency during photocharge process reaches 0.35% and the solar-to-output energy efficiency of the as-designed photoanode is 0.18%, a performance level previously achieved in hematite systems only with the assistance of external solar cells. Overall, the straightforward photoanode preparation process, the earth-abundant material choice along with the remarkable performance should be promising for the practical application of the solar rechargeable batteries. In addition, further advancements in the SRFB system can be anticipated, particularly in terms of microfluidic design to enhance mass transport and nanophotonic engineering to improve charge transfer and optical performance. These SRFB design concepts may open new avenues for reaching highly efficient solar redox flow batteries.

## Acknowledgment

The authors acknowledge the support of the Swiss National Science Foundation (Eccellenza Grant #194181). J.M. acknowledges the support of the China Scholarship Council (201906210091). The authors also acknowledge the support of the following experimental facilities at EPFL: Interdisciplinary Centre for Electron Microscopy (CIME) and Center of MicroNano Technology (CMI). Finally, the authors would like to thank Dr. Alan Bowman for his assistance with the discussion on absorption testing.

## Author Information

Corresponding Author

Giulia Tagliabue, Ecole Polytechnique Federale de Lausanne (EPFL), Lausanne, Switzerland; orcid.org/0000-0003-4587-728X; Email: giulia.tagliabue@epfl.ch

Other Authors

Jiaming Ma, Ecole Polytechnique Federale de Lausanne, Lausanne (EPFL), Switzerland;

orcid.org/ 0000-0003-4311-0640

Milad Sabzehparvar, Ecole Polytechnique Federale de Lausanne, Lausanne (EPFL), Switzerland;

orcid.org/ 0000-0001-5594-6889

Ziyan Pan, Ecole Polytechnique Federale de Lausanne, Lausanne (EPFL), Switzerland;

orcid.org/ 0009-0000-5585-5098

## Conflict of Interest

The authors declare no conflict of interest.

# SI-Nanostructured $Fe_2O_3/Cu_xO$ Heterojunction for Enhanced Solar Redox Flow Battery Performance

*Jiaming Ma[1], Milad Sabzehparvar[1], Ziyan Pan[1], Giulia Tagliabue[1]**

*[1] Laboratory of Nanoscience for Energy Technologies (LNET), STI, École Polytechnique Fédérale de Lausanne, 1015 Lausanne, Switzerland*

**Corresponding Author. Email: giulia.tagliabue@epfl.ch*

## Table of Contents

## Supporting Information 1 – Schematic of samples preparation

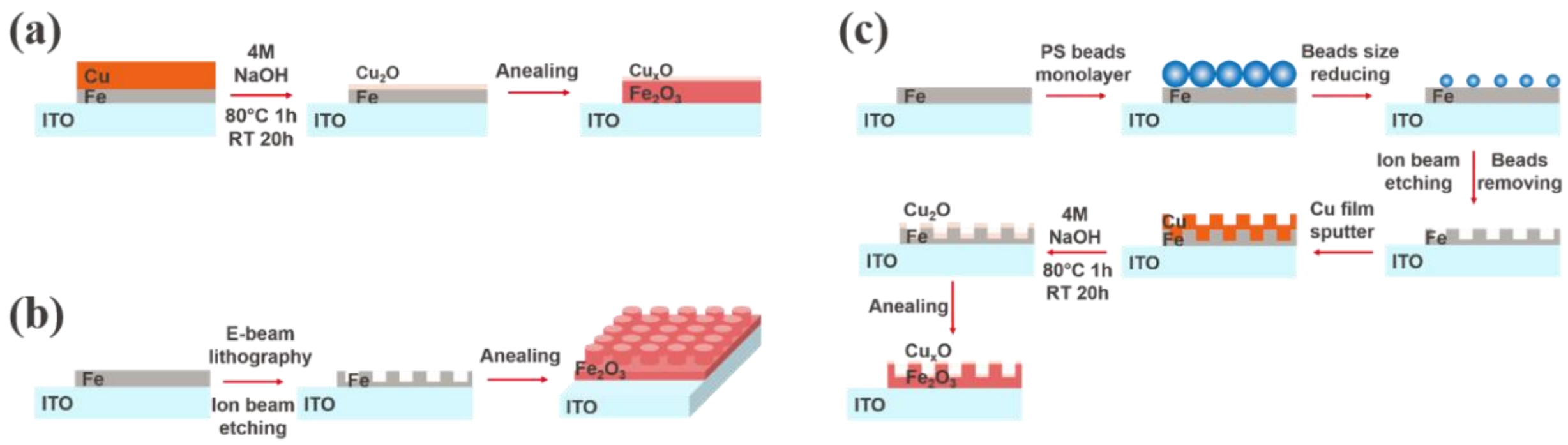

**Figure S1.** (a) Schematic of the synthesis of α-$Fe_2O_3$/$Cu_xO$ film photoanodes; (b) Schematic of the synthesis process of α-$Fe_2O_3$ nanopillar array; (c) Schematic of the synthesis of nanostructured α-$Fe_2O_3$/$Cu_xO$ photoanode. All the materials were sputtered on the ITO substrate.

**Table S1.** Summary of sample preparation process.

| Sample type | Sputtered Fe (nm) | Sputtered Cu (nm) | NaOH | Annealing | Sample |
|---|---|---|---|---|---|
| Planar $Fe_2O_3$ | 15 | 0 | NO | YES | 15F |
| | 30 | 0 | NO | YES | 30F |
| | 50 | 0 | NO | YES | 50F |
| Planar $Fe_2O_3$-CuO | 15 | 30 | NO | YES | 15/30 F-CuO |
| Planar $Fe_2O_3$-$Cu_xO$ | 15 | 30 | YES | YES | 15/30 F-$Cu_xO$ |
| | 30 | 30 | YES | YES | 30/30 F-$Cu_xO$ |
| | 50 | 30 | YES | YES | 50/30 F-$Cu_xO$ |
| Nanopatterned $Fe_2O_3$ | 30 | 0 | NO | YES | P100<br>P150<br>P200 |
| Nanostructured $Fe_2O_3$ | 30 | 0 | NO | YES | P |
| Nanostructured $Fe_2O_3$-$Cu_xO$ | 30 | 30 | YES | YES | P/$Cu_xO$ |

## Supporting Information 2 – Band alignment measurements

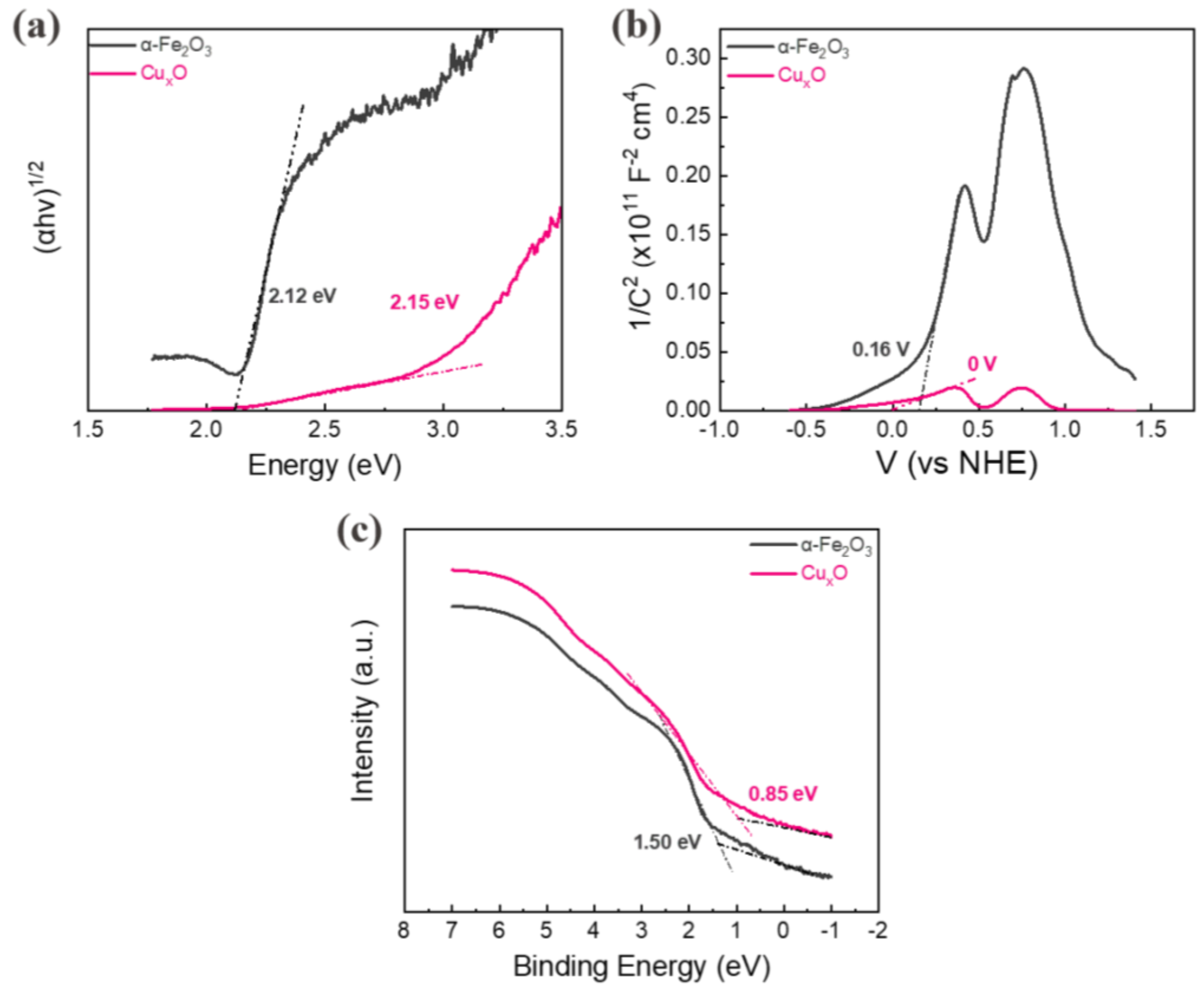

**Figure S2.** (a) Tauc plots, (b) Mott-schottky test and (c) UPS test of α-$Fe_2O_3$ and $Cu_xO$ to get bandgap, work function and valence band maxima (VBM); respectively.

The band gap value is estimated by identifying the intersection point between the linear portion of Tauc's plots and the X-axis. These plots of the absorption coefficient (α) were obtained using a UV-visible spectrophotometer (Shimadzu, UV-2600i) within the wavelength range of 300−800 nm. Mott−Schottky (MS) measurements were conducted under dark conditions using a three-electrode homemade cell via CHI 760E potentiostat. The photoanode served as the working electrode, while a Pt foil (99.99% purity, Alfa Aesar) acted as the counter electrode. Additionally, a leakage-free Ag/AgCl electrode (Innovative Instruments, Inc) was utilized as the reference electrode. The experiment was performed within a potential range of −0.8 to 1.2 V vs Ag/AgCl under anolyte, employing an applied frequency of 1 kHz to investigate the intrinsic electrical properties. The flat band potential ($F_{fb}$) value can be determined using the following equation[52]:

$$\mathrm{C}^{-2} = \frac{2}{\mathrm{q}\epsilon_0 \epsilon_s N_D}\left[V - F_{fb} - \frac{KT}{q}\right]$$

## Supporting Information 3 – Investigation of NaOH treatment time and Cu film thickness

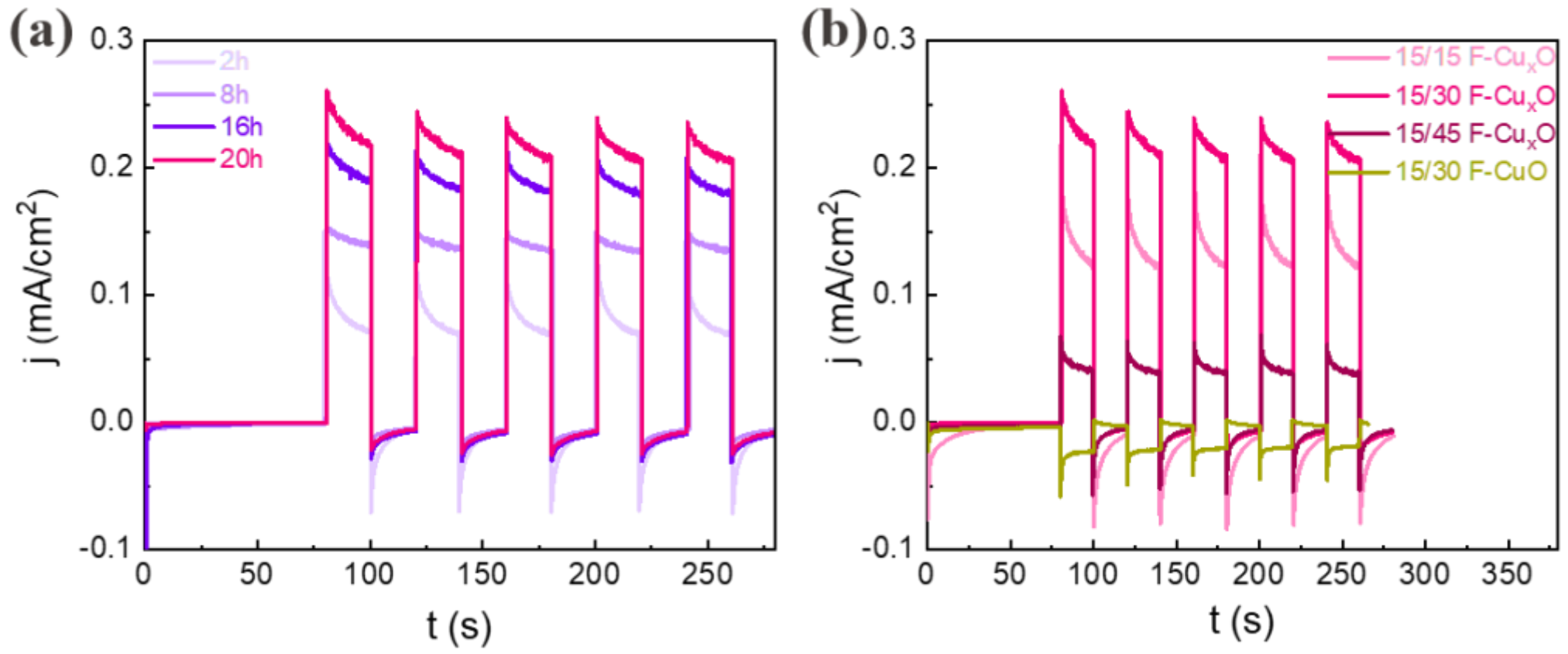

**Figure S3**. (a) Photoresponse behavior of 15/30 F-$Cu_xO$ with different NaOH treatment time; (b) Photocurrent performance of 15F-based heterojunctions with varying thicknesses of deposited Cu films subsequently treated for 20 h in NaOH.

## Supporting Information 4 – Scanning electron microscope of planer sample

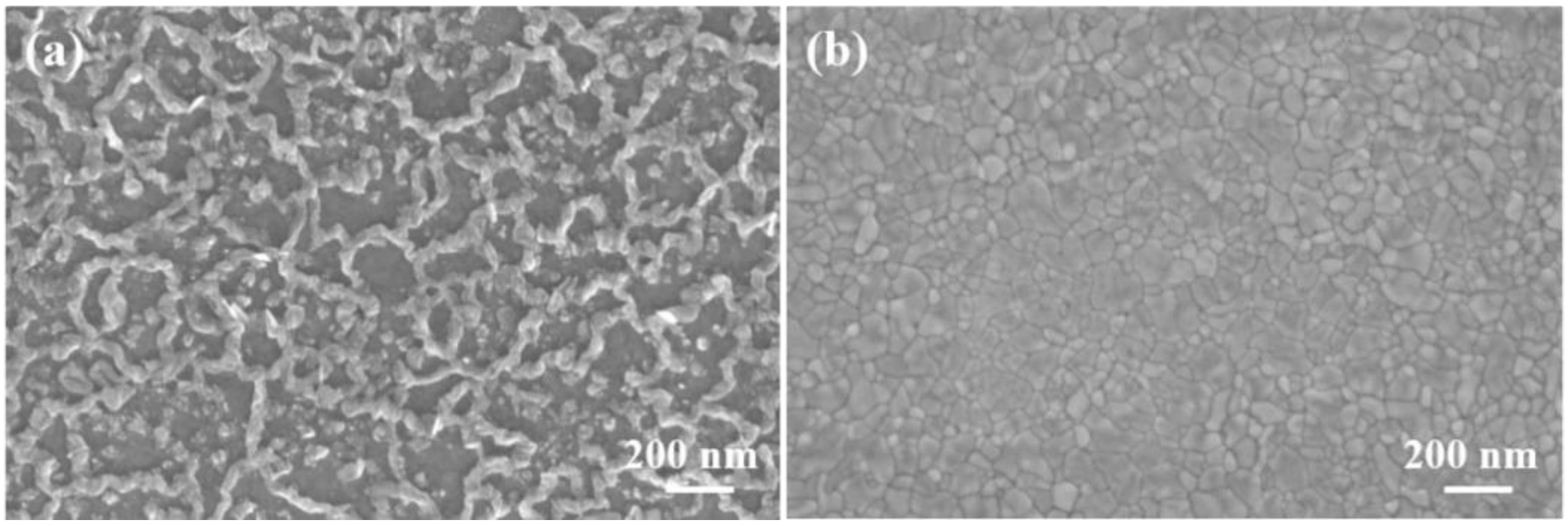

**Figure S4.** SEM of (a) 15F and (b) 15/30 F-$Cu_xO$.

## Supporting Information 5 – Stability of planer sample

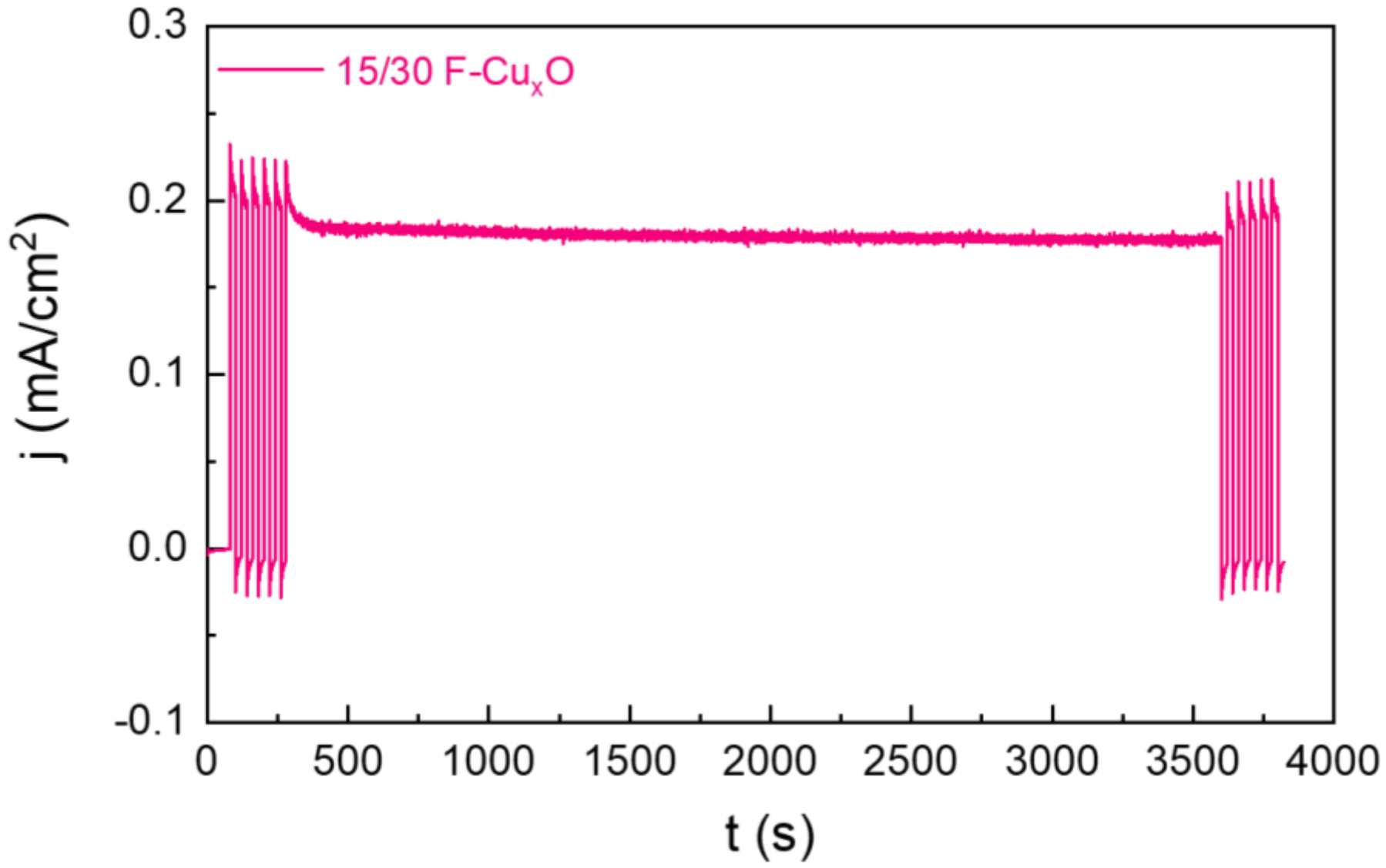

**Figure S1.** Stability test (1h) of 15/30 F-$Cu_xO$ with photoresponse before and after.

## Supporting Information 6 – Optical performance of hematite film with different thickness

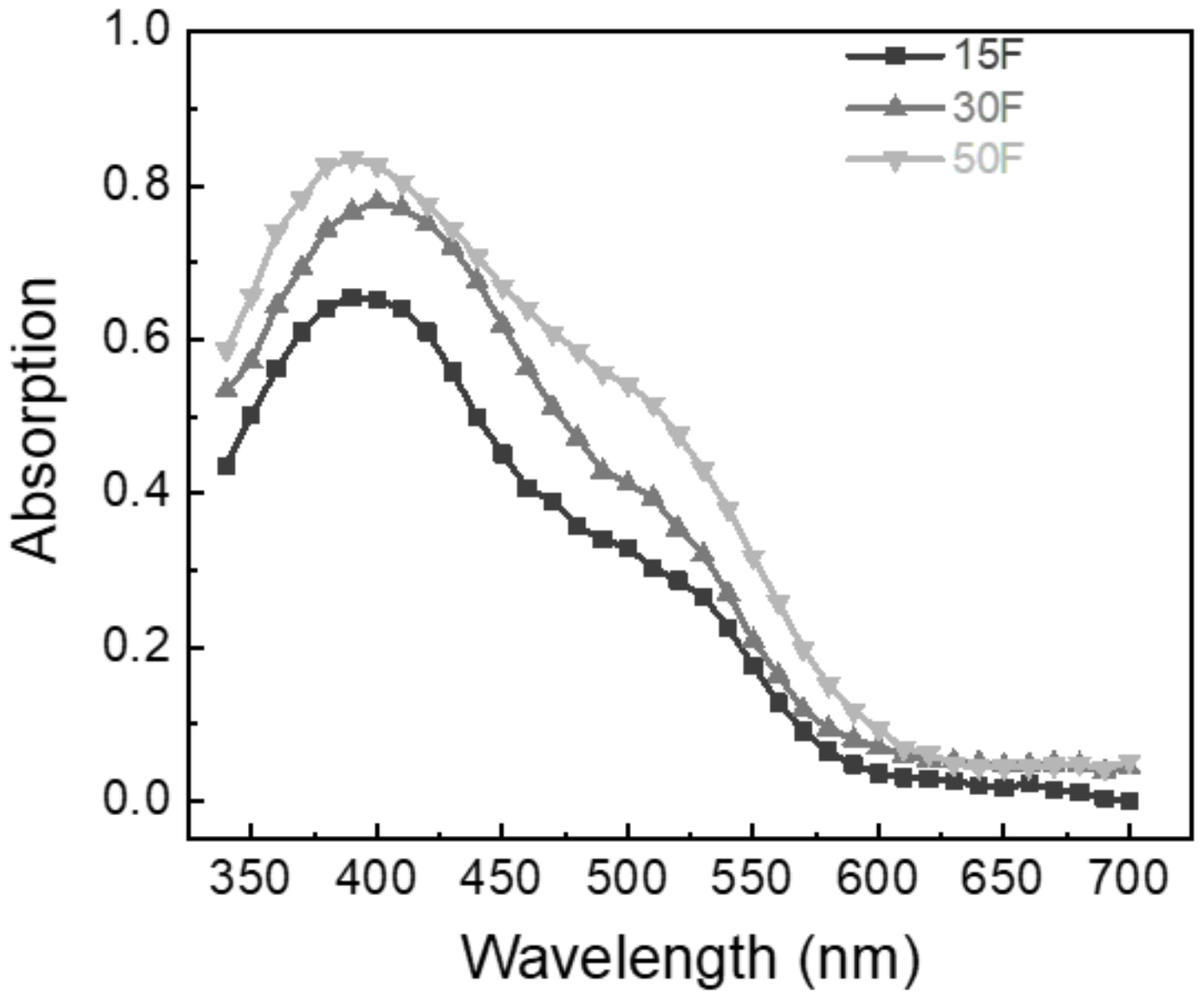

**Figure S6.** UV-Vis absorption spectrum of 15F, 30F and 50F.

## Supporting Information 7 – Optical simulation

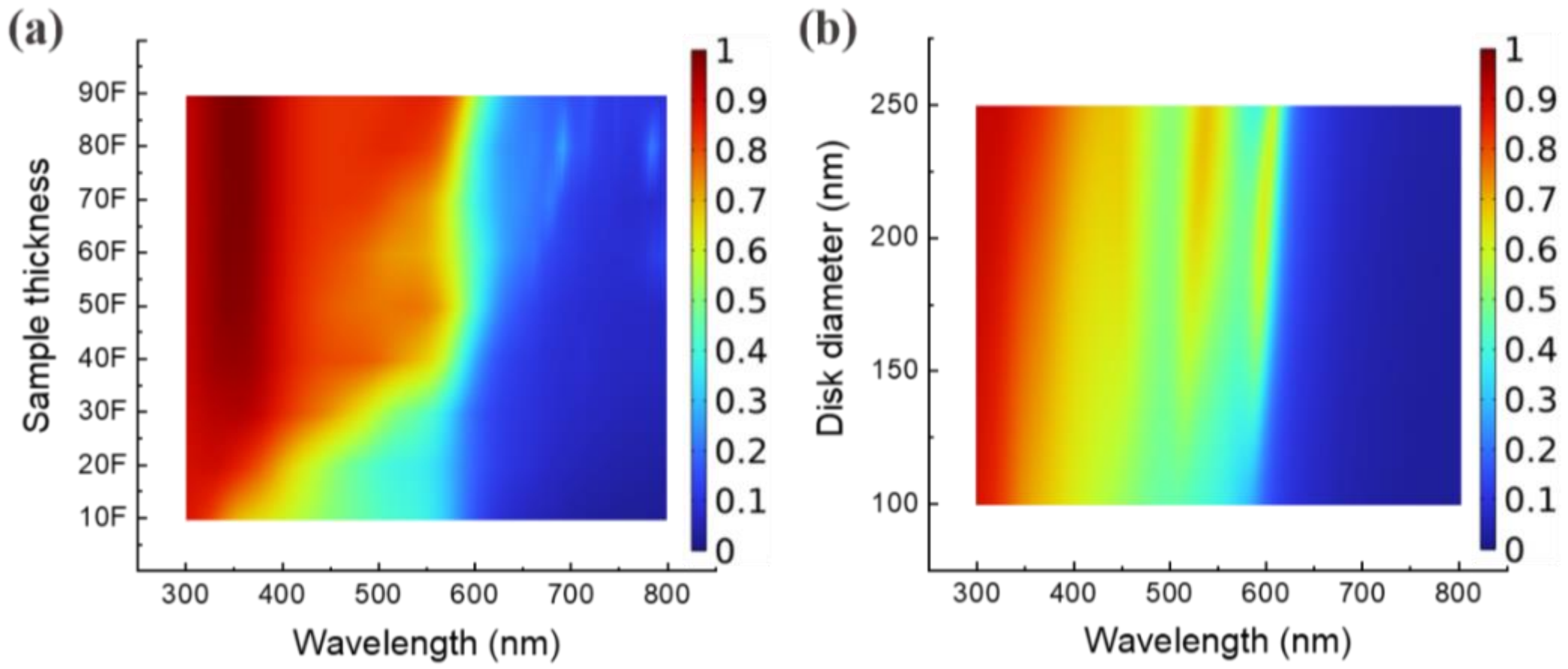

**Figure S2.** Simulated 2D absorption spectra of the (a) α-$Fe_2O_3$ film with different thickness and (b) 30F-based α-$Fe_2O_3$ nanopillar arrays with varying diameters under a periodicity of 300 nm.

## Supporting Information 8 – Photoresponse testing of various planer p-n junctions

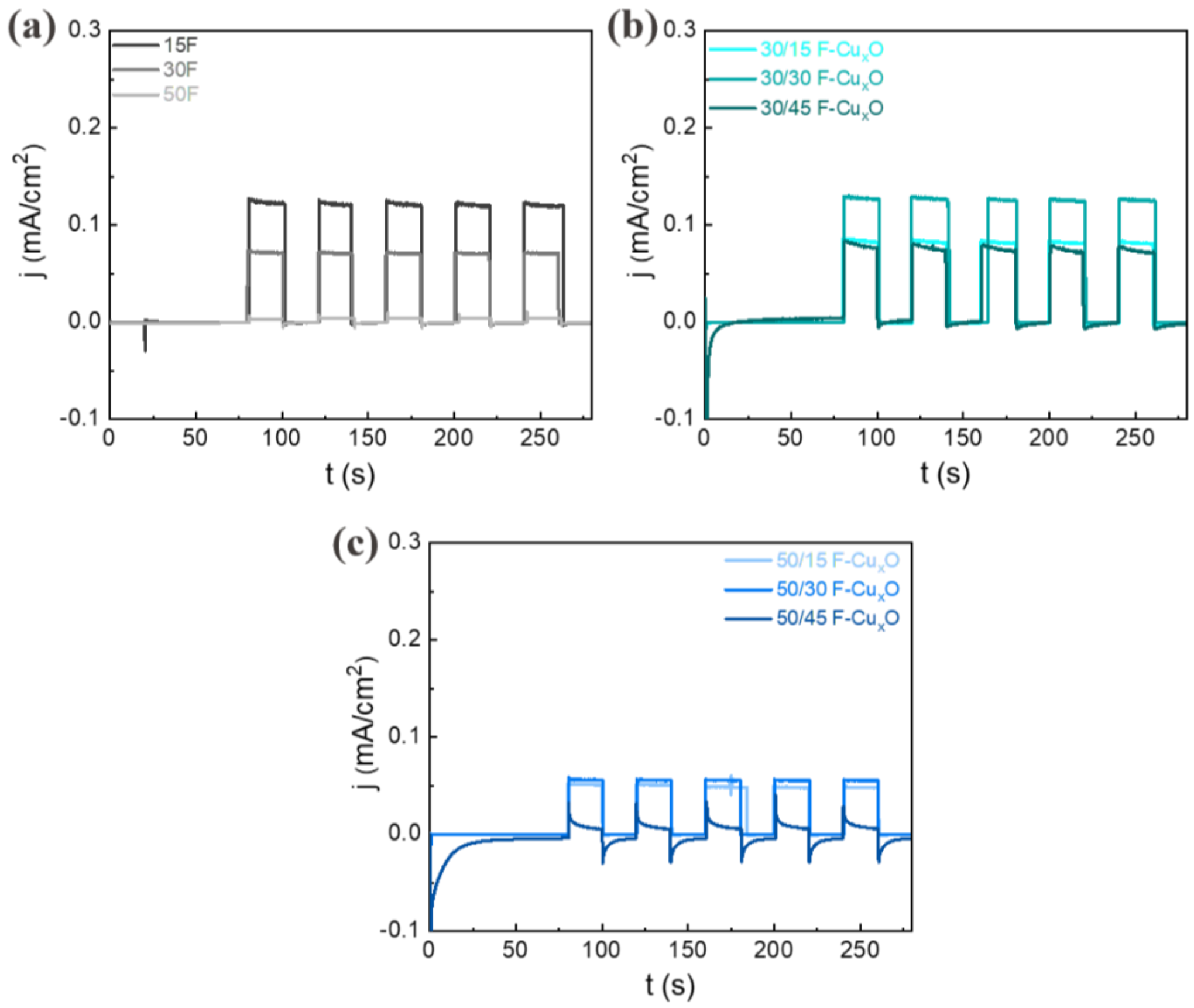

**Figure S3.** (a) The influence of α-$Fe_2O_3$ thickness on the photoelectrochemical performance; Photoresponse of (b) 30F-based and (c) 50F-based heterojunctions with varying thicknesses of deposited Cu films treated for 20 h in NaOH.

As a comparison, we measured the absorption spectra and photochemical activity of pure hematite films with different thicknesses (Figure S6 and FigureS8) under back illumination in ambient air conditions as described in Experimental Section. The measured absorption curves align well with numerical calculations (Figure S7a). The light absorption demonstrates a notable increase from 15F to 30F, followed by saturation upon reaching 50F. Yet, due to the limited charge mobility within hematite[53], electron-hole recombination significantly constrains the performance of the photoanodes. The photocurrent density decreases from 15F (0.12 mA/$cm^2$) to 30F (0.07 mA/$cm^2$) and experiences a significant decrease at 50F (0.01 mA/$cm^2$) thereafter (Figure S8a). Despite efforts to construct p-n junctions to improve charge transfer, the best samples obtained with thicker hematite films only exhibit a photocurrent density of 0.14 mA/$cm^2$ and 0.06 mA/$cm^2$ for 30/30 F-$Cu_xO$ and 50/30 F-$Cu_xO$, respectively (Figure S8b, c)

## Supporting Information 9 – Electrochemical impedance spectroscopy analysis

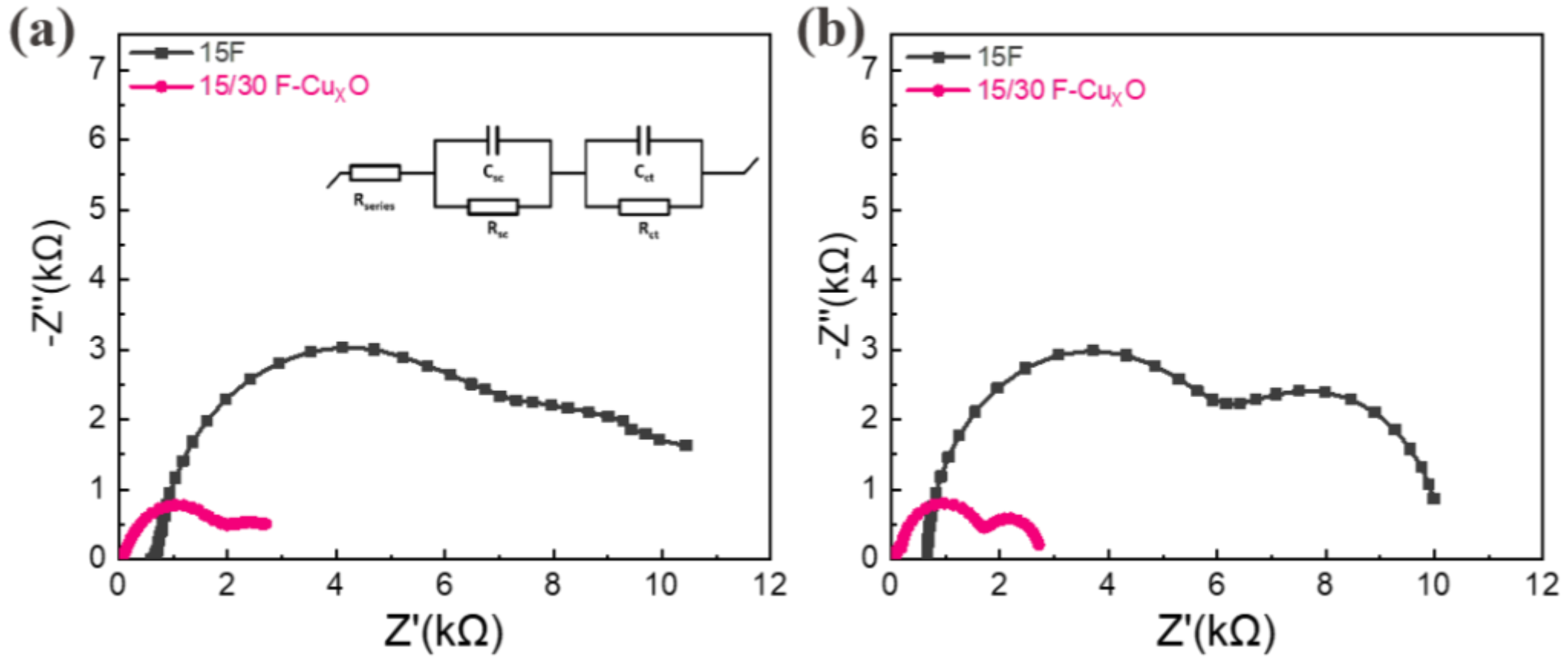

**Figure S9.** The (a) measured and (b) fitting EIS results of 15F and 15/30 F-$Cu_xO$. The inset is the equivalent electrical circuit.

Electrochemical Impedance Spectroscopy (EIS) analysis was performed with frequency ranging from 10 kHz to 10 mHz under 1 sun illumination to further investigate the mechanisms that occur at the photoanodes. The electrical circuit (the inset of Figure S9a) was used as a model to fit the PEC redox couple oxidation reaction process in the photoanodes. Where $R_{series}$ represents the series resistance, which includes the resistance at the ITO/photoanode interface, the ionic conductivity in the electrolyte and the external contact resistance; $R_{sc}$ and $C_{sc}$ are models of charge transfer behavior in the internal semiconductor, while $R_{ct}$ and $C_{ct}$ are the analogous reactions that take place at the semiconductor/electrolyte interface. The fitting Nyquist plots are shown in Figure S9b, which is similar to the measured EIS results.

## Supporting Information 10 – SEM and optical microscopy image of different patterns

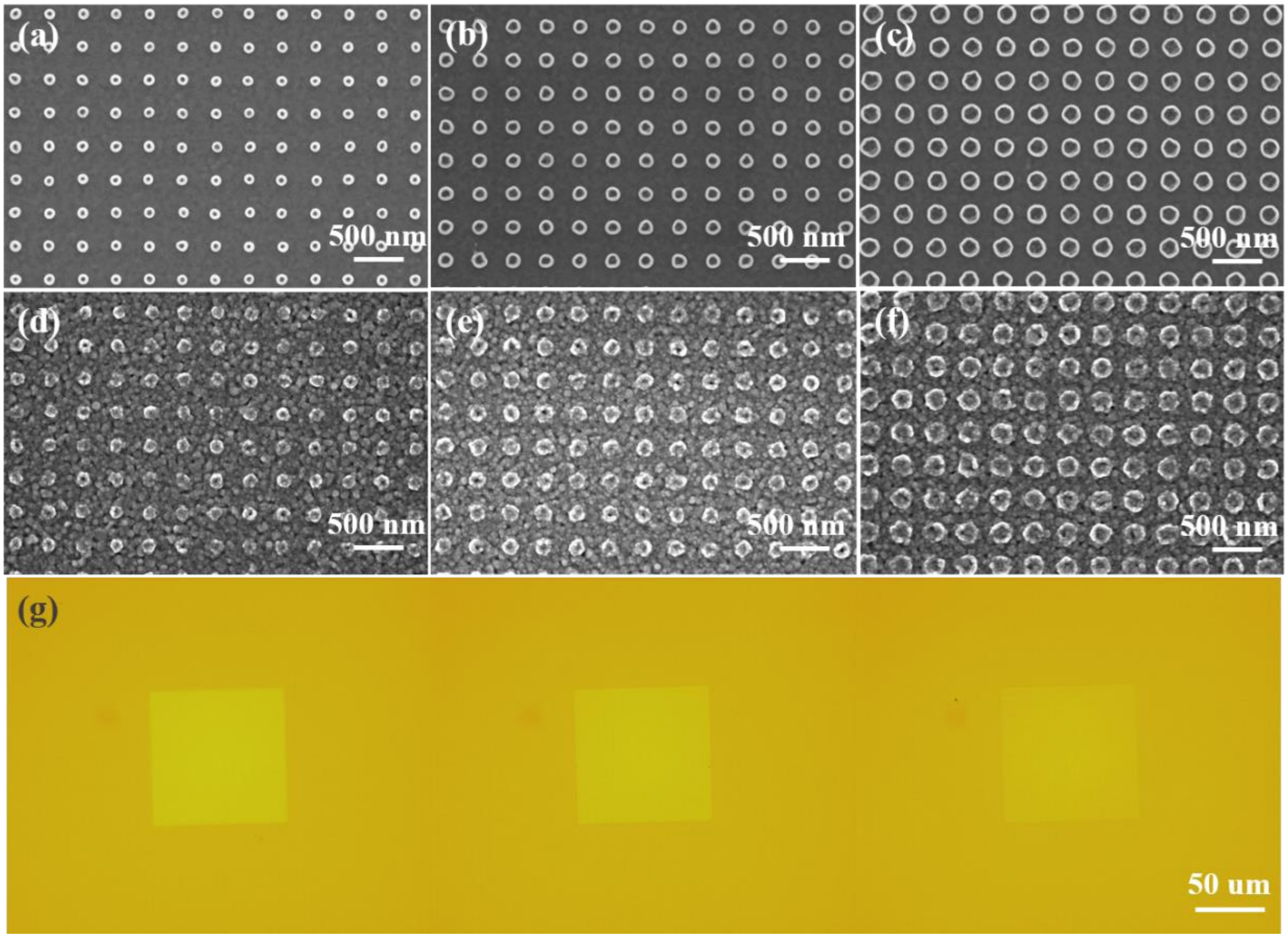

**Figure S10.** SEM of (a, d) P100, (b, e) P150 and (c, f) P200 before and after annealing; (g) Optical microscopy image of the α-$Fe_2O_3$ nanopillar arrays.

The corresponding optical, white-light transmission images (Figure S10g) exhibit the different optical appearances of the three patterns, providing a direct way to observe the tunability of the optical properties with varying diameters.

## Supporting Information 11 – Microscale absorption measurements

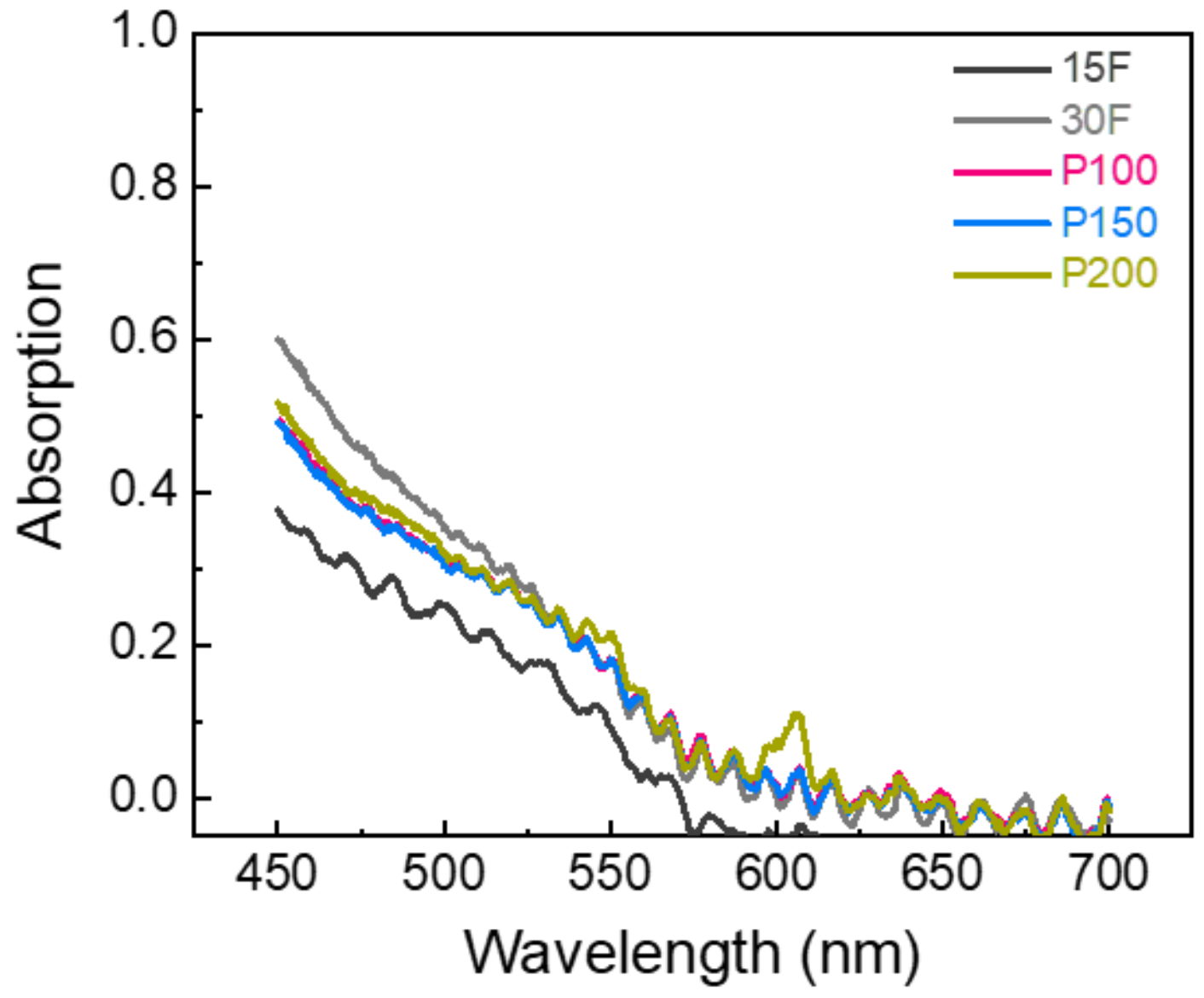

**Figure S11.** Microscale absorption measurements of the fabricated α-$Fe_2O_3$ nanopillar arrays as well as α-$Fe_2O_3$ films.

The absorption depth for incident photons of hematite[54] is 40-100 nm for wavelengths ranging from 450 to 550 nm. A significant enhancement in broadband light absorption can be observed from Figure S11 as the thickness of the thin film sample increases from 15F (~30 nm hematite) to 30F (~60 nm hematite). However, the hole diffusion length limitation for hematite[55] necessitated a sacrifice in the thickness of the photoanode. Nanoengineering could potentially serve as a solution strategy to balance optical performance and charge carrier transport. It is worth noting that all the 30F-based nanopillar arrays (even those with only a 10 nm thickness of continuous α-$Fe_2O_3$ thin film underneath) exhibit comparable optical absorption to the 30F sample, significantly stronger than the 15F sample. This enhancement could be attributed to the light-trapping effect of the nanopillar array[56]. Additionally, P200 enables sunlight to induce optical resonances at 550 nm and 600 nm, consistent with our simulation findings (Figure S7b). This phenomenon can enhance the light intensity within the photoanode[57].

## Supporting Information 12 – Tip approaching curve

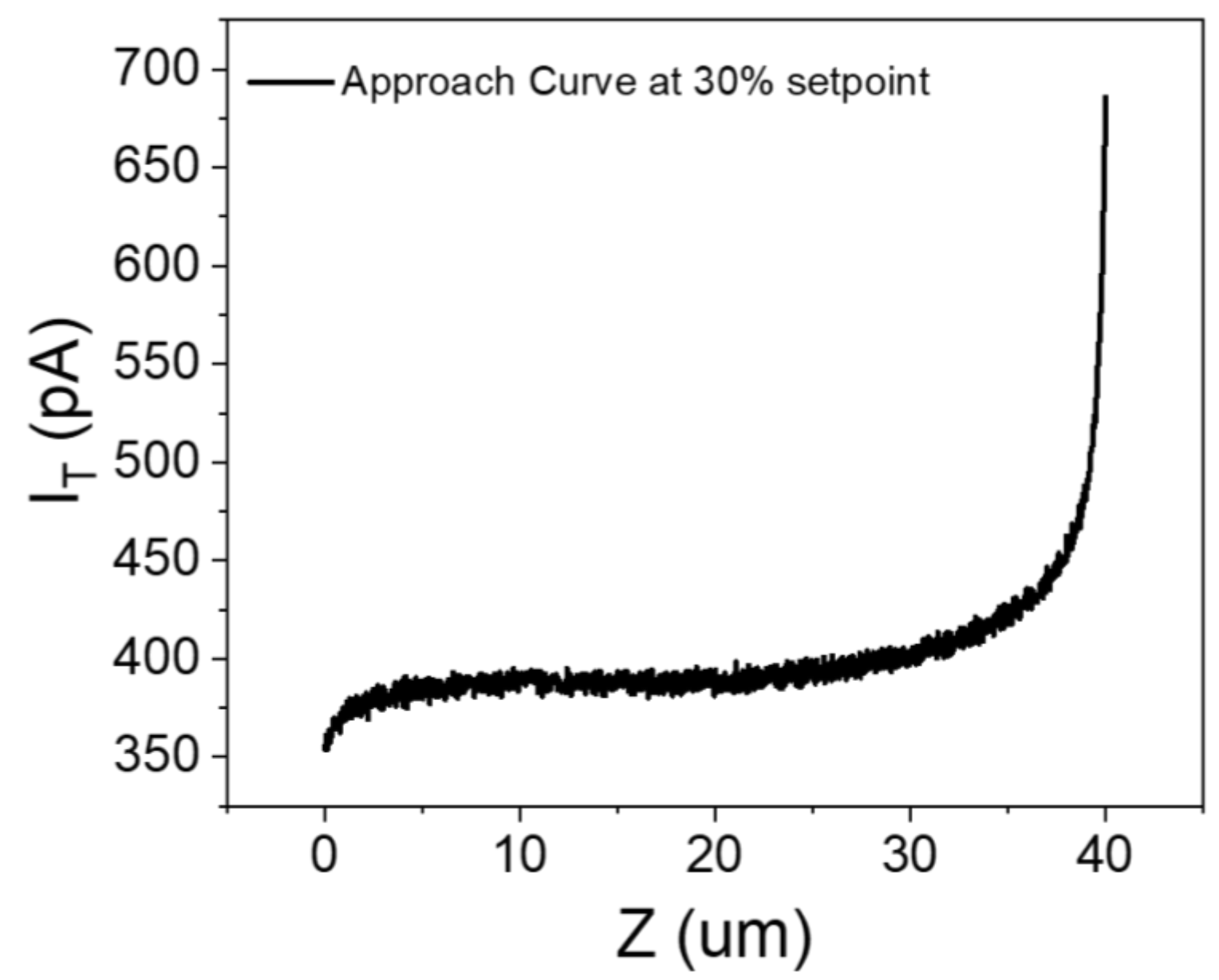

**Figure S12.** Tip approach curve. Experimental current–distance curves (points) obtained with the tip approaching the hematite film part of the substrate. The solution contained 4 mM $Fe(CN)_6^{4-}$ and 0.4 M NaOH. Tip potential $E_T$=0.4 V vs. Ag/AgCl and an unbiased grounded substrate in dark.

## Supporting Information 13 – SEM for nanostructures

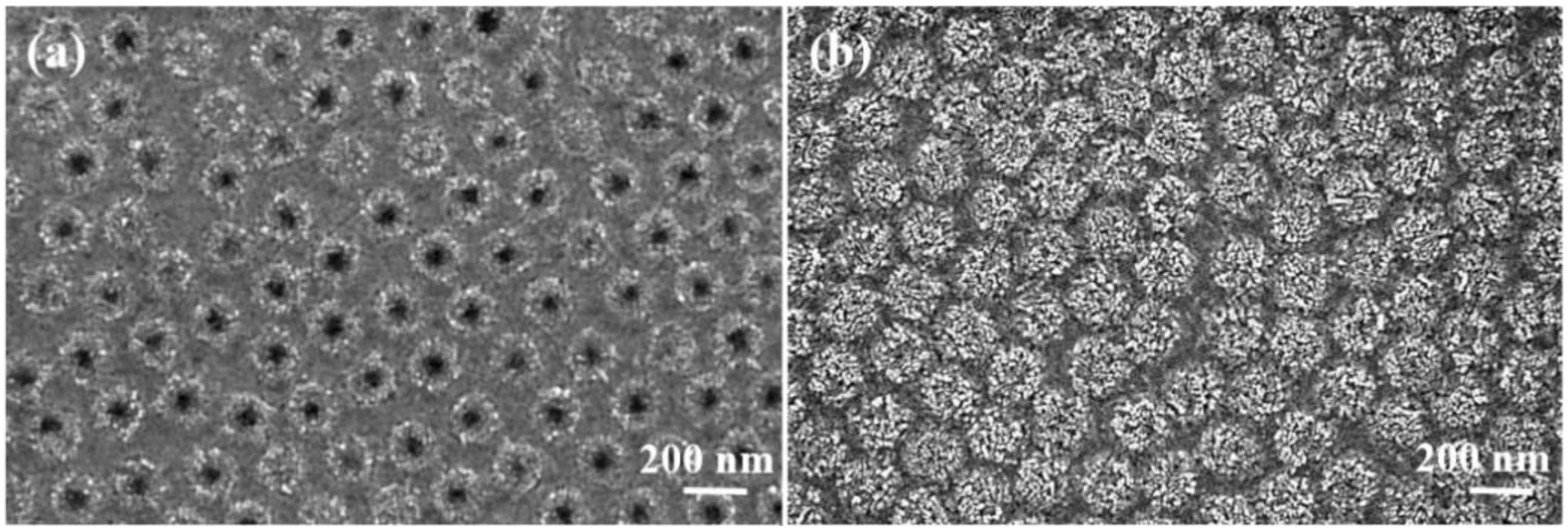

**Figure 13.** SEM of (a) P and (b) P/$Cu_xO$.

## Supporting Information 14 – Photoresponse and stability for nanostructures

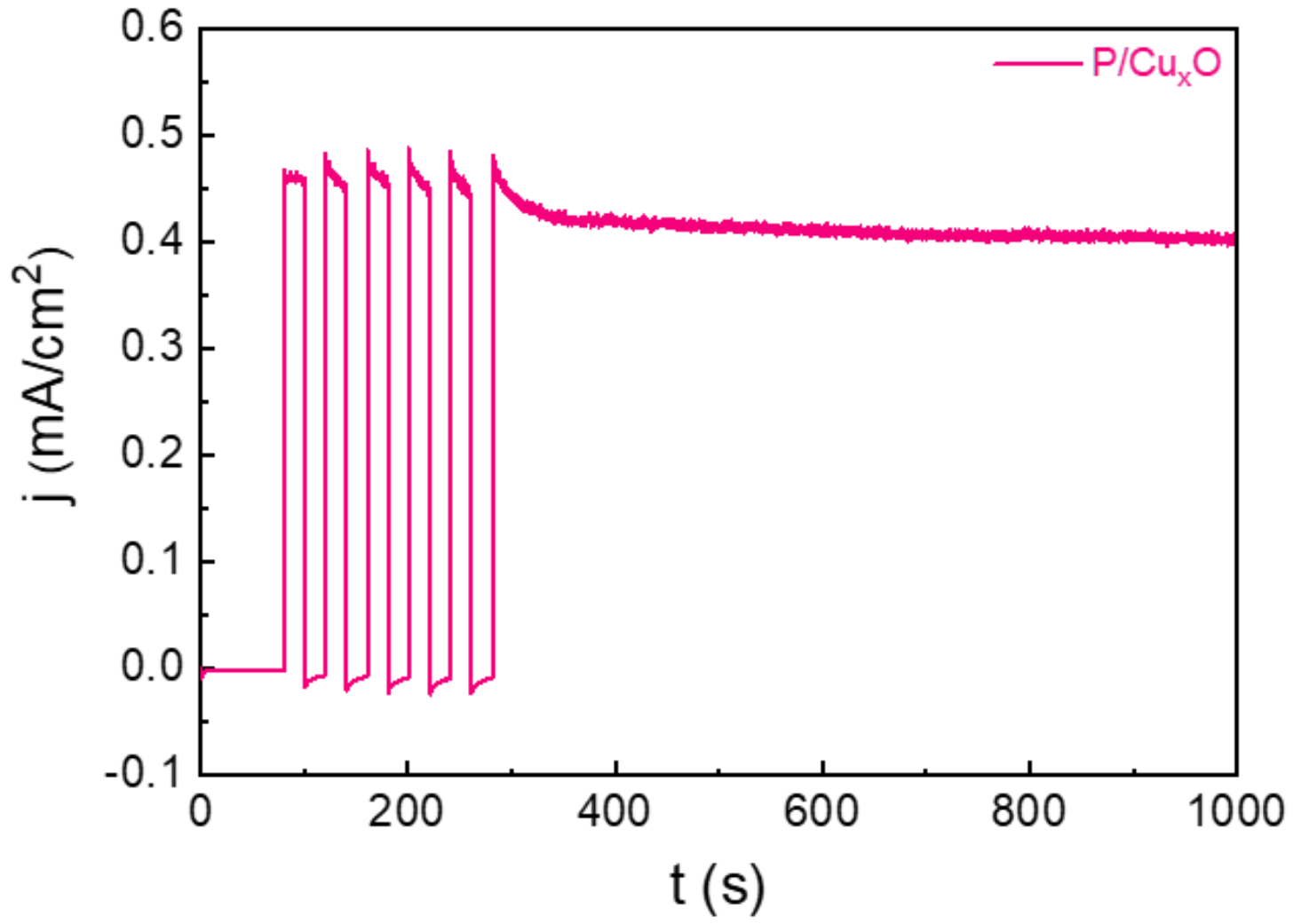

**Figure S4.** Stability of j-t for P/$Cu_xO$.

## Supporting Information 15 – EIS analysis for nanostructures

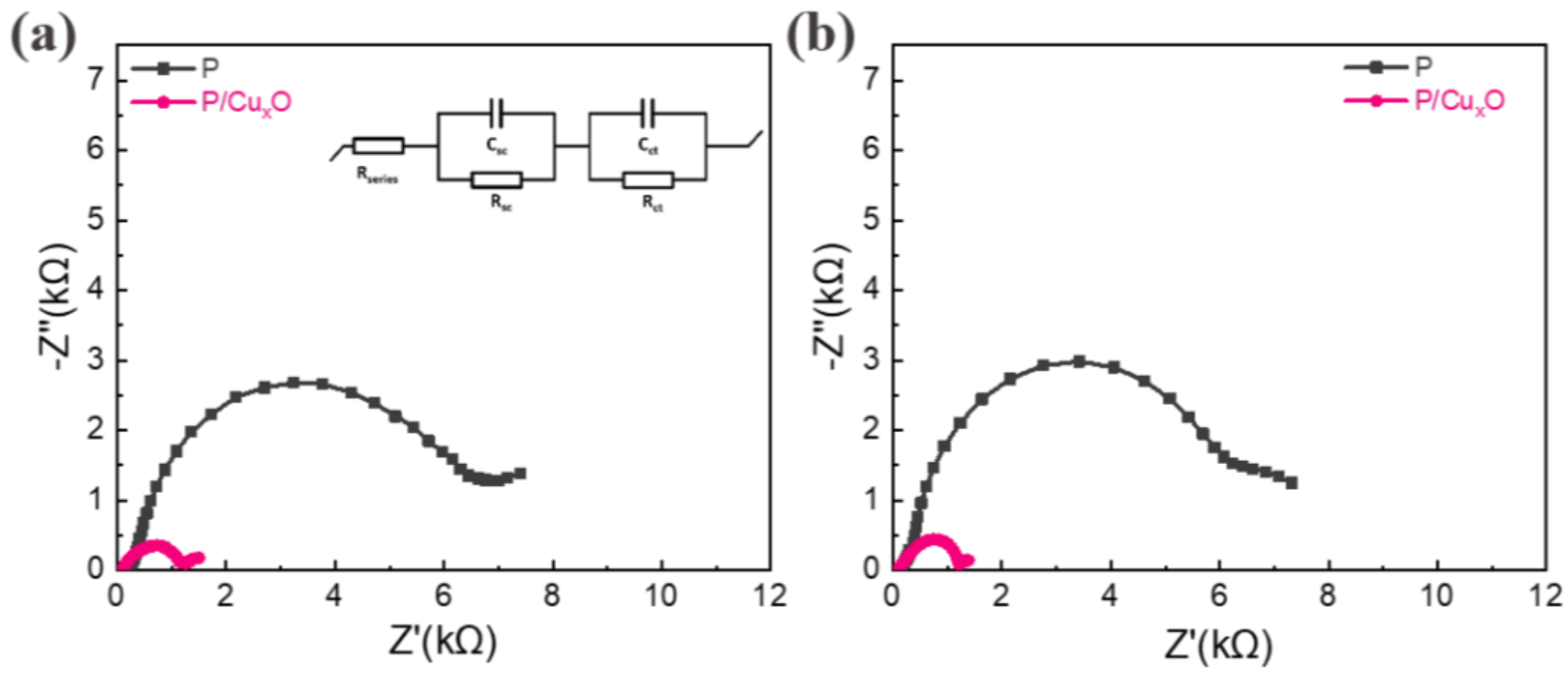

**Figure S5.** The (a) measured and (b) fitting EIS results of α-$Fe_2O_3$ P and P/$Cu_xO$. The inset is the equivalent electrical circuit.

By using the same equivalent electrical circuit as Figure S9, we can get the $R_{series}$, $R_{sc}$ , $R_{ct}$ and $C_{sc}$ for α-$Fe_2O_3$ P is 160 Ω, 4075 Ω, 2028 Ω and 0.03×$10^{-3}$ F, respectively. The $R_{series}$, $R_{sc}$ , $R_{ct}$ and $C_{sc}$ for P/$Cu_xO$ is 90 Ω, 500 Ω, 757 Ω and 1.4×$10^{-3}$ F, respectively.

## Reference

(1) Fu, H.-C.; Li, W.; Yang, Y.; Lin, C.-H.; Veyssal, A.; He, J.-H.; Jin, S. An Efficient and Stable Solar Flow Battery Enabled by a Single-Junction GaAs Photoelectrode. Nat. Commun. 2021, 12 (1), 156. https://doi.org/10.1038/s41467-020-20287-w.

(2) Wedege, K.; Bae, D.; Dražević, E.; Mendes, A.; Vesborg, P. C. K.; Bentien, A. Unbiased, Complete Solar Charging of a Neutral Flow Battery by a Single Si Photocathode. *RSC Adv.* **2018**, *8* (12), 6331–6340. https://doi.org/10.1039/C8RA00319J.

(3) Feng, H.; Liu, D.; Zhang, Y.; Shi, X.; Esan, O. C.; Li, Q.; Chen, R.; An, L. Advances and Challenges in Photoelectrochemical Redox Batteries for Solar Energy Conversion and Storage. *Adv. Energy Mater.* **2022**, *12* (24), 2200469. https://doi.org/10.1002/aenm.202200469.

(4) Cao, L.; Skyllas-Kazacos, M.; Wang, D.-W. Solar Redox Flow Batteries: Mechanism, Design, and Measurement. *Adv. Sustain. Syst.* **2018**, *2* (8–9), 1800031. https://doi.org/10.1002/adsu.201800031.

(5) da Silva Lopes, T.; Dias, P.; Monteiro, R.; Vilanova, A.; Ivanou, D.; Mendes, A. A 25 Cm2 Solar Redox Flow Cell: Facing the Engineering Challenges of Upscaling. *Adv. Energy Mater.* **2022**, *12* (5), 2102893. https://doi.org/10.1002/aenm.202102893.

(6) Li, J.; Cushing, S. K.; Zheng, P.; Meng, F.; Chu, D.; Wu, N. Plasmon-Induced Photonic and Energy-Transfer Enhancement of Solar Water Splitting by a Hematite Nanorod Array. *Nat. Commun.* **2013**, *4* (1), 2651. https://doi.org/10.1038/ncomms3651.

(7) Yun, S.; Qin, Y.; R. Uhl, A.; Vlachopoulos, N.; Yin, M.; Li, D.; Han, X.; Hagfeldt, A. New-Generation Integrated Devices Based on Dye-Sensitized and Perovskite Solar Cells. *Energy Environ. Sci.* **2018**, *11* (3), 476–526. https://doi.org/10.1039/C7EE03165C.

(8) Bae, D.; Faasse, G. M.; Kanellos, G.; Smith, W. A. Unravelling the Practical Solar Charging Performance Limits of Redox Flow Batteries Based on a Single Photon Device System. *Sustain. Energy Fuels* **2019**, *3* (9), 2399–2408. https://doi.org/10.1039/C9SE00333A.

(9) McKone, J. R.; DiSalvo, F. J.; Abruña, H. D. Solar Energy Conversion, Storage, and Release Using an Integrated Solar-Driven Redox Flow Battery. *J. Mater. Chem. A* **2017**, *5* (11), 5362–5372. https://doi.org/10.1039/C7TA00555E.

(10) Bae, D.; Kanellos, G.; Faasse, G. M.; Dražević, E.; Venugopal, A.; Smith, W. A. Design Principles for Efficient Photoelectrodes in Solar Rechargeable Redox Flow Cell Applications. *Commun. Mater.* **2020**, *1* (1), 1–9. https://doi.org/10.1038/s43246-020-0020-7.

(11) Liao, S.; Zong, X.; Seger, B.; Pedersen, T.; Yao, T.; Ding, C.; Shi, J.; Chen, J.; Li, C. Integrating a Dual-Silicon Photoelectrochemical Cell into a Redox Flow Battery for Unassisted Photocharging. *Nat. Commun.* **2016**, *7* (1), 11474. https://doi.org/10.1038/ncomms11474.

(12) Li, W.; Zheng, J.; Hu, B.; Fu, H.-C.; Hu, M.; Veyssal, A.; Zhao, Y.; He, J.-H.; Liu, T. L.; Ho-Baillie, A.; Jin,
S. High-Performance Solar Flow Battery Powered by a Perovskite/Silicon Tandem Solar Cell. *Nat. Mater.* **2020**, *19* (12), 1326–1331. https://doi.org/10.1038/s41563-020-0720-x.

(13) Li, W.; Fu, H.-C.; Zhao, Y.; He, J.-H.; Jin, S. 14.1% Efficient Monolithically Integrated Solar Flow Battery. *Chem* **2018**, *4* (11), 2644–2657. https://doi.org/10.1016/j.chempr.2018.08.023.

(14) Li, W.; Jin, S. Design Principles and Developments of Integrated Solar Flow Batteries. *Acc. Chem. Res.* **2020**, *53* (11), 2611–2621. https://doi.org/10.1021/acs.accounts.0c00373.

(15) Lin, Y.; Yuan, G.; Sheehan, S.; Zhou, S.; Wang, D. Hematite -Based Solar Water Splitting: Challenges and Opportunities. *Energy Environ. Sci.* **2011**, *4* (12), 4862–4869. https://doi.org/10.1039/C1EE01850G.

(16) Khataee, A.; Azevedo, J.; Dias, P.; Ivanou, D.; Dražević, E.; Bentien, A.; Mendes, A. Integrated Design of Hematite and Dye-Sensitized Solar Cell for Unbiased Solar Charging of an Organic-Inorganic Redox Flow Battery. *Nano Energy* **2019**, *62*, 832–843. https://doi.org/10.1016/j.nanoen.2019.06.001.
(17) Tolod, K. R.; Hernández, S.; Quadrelli, E. A.; Russo, N. Chapter 4 - Visible Light-Driven Catalysts for Water Oxidation: Towards Solar Fuel Biorefineries. In *Studies in Surface Science and Catalysis*; Albonetti, S., Perathoner, S., Quadrelli, E. A., Eds.; Horizons in Sustainable Industrial Chemistry and Catalysis; Elsevier, 2019; Vol. 178, pp 65–84. https://doi.org/10.1016/B978-0-444-64127-4.00004-5.
(18) Gao, R.-T.; Zhang, J.; Nakajima, T.; He, J.; Liu, X.; Zhang, X.; Wang, L.; Wu, L. Single-Atomic-Site Platinum Steers Photogenerated Charge Carrier Lifetime of Hematite Nanoflakes for Photoelectrochemical Water Splitting. *Nat. Commun.* **2023**, *14* (1), 2640. https://doi.org/10.1038/s41467-023-38343-6.
(19) Piekner, Y.; S. Ellis, D.; A. Grave, D.; Tsyganok, A.; Rothschild, A. Wasted Photons: Photogeneration Yield and Charge Carrier Collection Efficiency of Hematite Photoanodes for Photoelectrochemical Water Splitting. *Energy Environ. Sci.* **2021**, *14* (8), 4584–4598. https://doi.org/10.1039/D1EE01772A.
(20) Le Formal, F.; Grätzel, M.; Sivula, K. Controlling Photoactivity in Ultrathin Hematite Films for Solar Water-Splitting. *Adv. Funct. Mater.* **2010**, *20* (7), 1099–1107. https://doi.org/10.1002/adfm.200902060.
(21) Gardner, R. F. G.; Sweett, F.; Tanner, D. W. The Electrical Properties of Alpha Ferric Oxide—II.: Ferric Oxide of High Purity. *J. Phys. Chem. Solids* **1963**, *24* (10), 1183–1196. https://doi.org/10.1016/0022-3697(63)90235-X.
(22) Gao, R.; Yan, D. Recent Development of Ni/Fe-Based Micro/Nanostructures toward Photo/Electrochemical Water Oxidation. *Adv. Energy Mater.* **2020**, *10* (11), 1900954. https://doi.org/10.1002/aenm.201900954.
(23) Tang, R.; Zhou, S.; Zhang, Z.; Zheng, R.; Huang, J. Engineering Nanostructure–Interface of Photoanode Materials Toward Photoelectrochemical Water Oxidation. *Adv. Mater.* **2021**, *33* (17), 2005389. https://doi.org/10.1002/adma.202005389.
(24) Lee, J.-W.; Cho, K.-H.; Yoon, J.-S.; Kim, Y.-M.; Sung, Y.-M. Photoelectrochemical Water Splitting Using One-Dimensional Nanostructures. *J. Mater. Chem. A* **2021**, *9* (38), 21576–21606. https://doi.org/10.1039/D1TA04829E.
(25) Krause, L.; Skibińska, K.; Rox, H.; Baumann, R.; Marzec, M. M.; Yang, X.; Mutschke, G.; Żabiński, P.; Lasagni, A. F.; Eckert, K. Hydrogen Bubble Size Distribution on Nanostructured Ni Surfaces: Electrochemically Active Surface Area Versus Wettability. *ACS Appl. Mater. Interfaces* **2023**, *15* (14), 18290–18299. https://doi.org/10.1021/acsami.2c22231.
(26) Kiani, F.; Sterl, F.; Tsoulos, T. V.; Weber, K.; Giessen, H.; Tagliabue, G. Ultra-Broadband and Omnidirectional Perfect Absorber Based on Copper Nanowire/Carbon Nanotube Hierarchical Structure. *ACS Photonics* **2020**, *7* (2), 366–374. https://doi.org/10.1021/acsphotonics.9b01658.
(27) Kim, S. J.; Thomann, I.; Park, J.; Kang, J.-H.; Vasudev, A. P.; Brongersma, M. L. Light Trapping for Solar Fuel Generation with Mie Resonances. *Nano Lett.* **2014**, *14* (3), 1446–1452. https://doi.org/10.1021/nl404575e.
(28) Li, F.; Li, J.; Zhang, J.; Gao, L.; Long, X.; Hu, Y.; Li, S.; Jin, J.; Ma, J. NiO Nanoparticles Anchored on Phosphorus-Doped α-Fe2O3 Nanoarrays: An Efficient Hole Extraction p–n Heterojunction Photoanode for Water Oxidation. *ChemSusChem* **2018**, *11* (13), 2156–2164. https://doi.org/10.1002/cssc.201800571.
(29) Low, J.; Yu, J.; Jaroniec, M.; Wageh, S.; Al-Ghamdi, A. A. Heterojunction Photocatalysts. *Adv. Mater.* **2017**, *29* (20), 1601694. https://doi.org/10.1002/adma.201601694.
(30) Zhang, H.; Zhang, P.; Zhao, J.; Liu, Y.; Huang, Y.; Huang, H.; Yang, C.; Zhao, Y.; Wu, K.; Fu, X.; Jin, S.; Hou, Y.; Ding, Z.; Yuan, R.; Roeffaers, M. B. J.; Zhong, S.; Long, J. The Hole-Tunneling Heterojunction of Hematite-Based Photoanodes Accelerates Photosynthetic Reaction. *Angew. Chem. Int. Ed.* **2021**, *60* (29), 16009–16018. https://doi.org/10.1002/anie.202102983.

(31) Kodan, N.; Agarwal, K.; Mehta, B. R. All-Oxide α-Fe2O3/H:TiO2 Heterojunction Photoanode: A Platform for Stable and Enhanced Photoelectrochemical Performance through Favorable Band Edge Alignment. *J. Phys. Chem. C* **2019**, *123* (6), 3326–3335. https://doi.org/10.1021/acs.jpcc.8b10794.

(32) Barreca, D.; Carraro, G.; Gasparotto, A.; Maccato, C.; Warwick, M. E. A.; Kaunisto, K.; Sada, C.; Turner, S.; Gönüllü, Y.; Ruoko, T.-P.; Borgese, L.; Bontempi, E.; Van Tendeloo, G.; Lemmetyinen, H.; Mathur, S. Fe2O3–TiO2 Nano-Heterostructure Photoanodes for Highly Efficient Solar Water Oxidation. *Adv. Mater. Interfaces* **2015**, *2* (17), 1500313. https://doi.org/10.1002/admi.201500313.

(33) Zhang, P.; Yu, L.; Lou, X. W. (David). Construction of Heterostructured Fe2O3-TiO2 Microdumbbells for Photoelectrochemical Water Oxidation. *Angew. Chem. Int. Ed.* **2018**, *57* (46), 15076–15080. https://doi.org/10.1002/anie.201808104.

(34) You Zheng, J.; Van, T.-K.; U. Pawar, A.; Woo Kim, C.; Soo Kang, Y. One-Step Transformation of Cu to Cu 2 O in Alkaline Solution. *RSC Adv.* **2014**, *4* (36), 18616–18620. https://doi.org/10.1039/C4RA01174K.

(35) Thangamuthu, M.; Santschi, C.; Martin, O. J. F. Reliable Langmuir Blodgett Colloidal Masks for Large Area Nanostructure Realization. *Thin Solid Films* **2020**, *709*, 138195. https://doi.org/10.1016/j.tsf.2020.138195.

(36) Mireles, M.; Soule, C. W.; Dehghani, M.; Gaborski, T. R. Use of Nanosphere Self-Assembly to Pattern Nanoporous Membranes for the Study of Extracellular Vesicles. *Nanoscale Adv.* **2020**, *2* (10), 4427–4436. https://doi.org/10.1039/D0NA00142B.

(37) Ma, J.; Oh, K.; Tagliabue, G. Understanding Wavelength-Dependent Synergies between Morphology and Photonic Design in TiO2-Based Solar Powered Redox Cells. *J. Phys. Chem. C* **2023**, *127* (1), 11–21. https://doi.org/10.1021/acs.jpcc.2c05893.

(38) Kiani, F.; Bowman, A. R.; Sabzehparvar, M.; Karaman, C. O.; Sundararaman, R.; Tagliabue, G. Transport and Interfacial Injection of D-Band Hot Holes Control Plasmonic Chemistry. *ACS Energy Lett.* **2023**, *8* (10), 4242–4250. https://doi.org/10.1021/acsenergylett.3c01505.

(39) Bowman, A. R.; Ma, J.; Kiani, F.; García Martínez, G.; Tagliabue, G. Best Practices in Measuring Absorption at the Macro- and Microscale. *APL Photonics* **2024**, *9* (6), 061101. https://doi.org/10.1063/5.0210830.

(40) *Optical Constants*. https://apps.dtic.mil/sti/citations/ADA158623 (accessed 2024-04-18).

(41) König, T. A. F.; Ledin, P. A.; Kerszulis, J.; Mahmoud, Mahmoud. A.; El-Sayed, M. A.; Reynolds, J. R.; Tsukruk, V. V. Electrically Tunable Plasmonic Behavior of Nanocube–Polymer Nanomaterials Induced by a Redox-Active Electrochromic Polymer. *ACS Nano* **2014**, *8* (6), 6182–6192. https://doi.org/10.1021/nn501601e.

(42) Saleem, M.; Al-Kuhaili, M. F.; Durrani, S. M. A.; Bakhtiari, I. A. Characterization of Nanocrystalline α-Fe2O3 Thin Films Grown by Reactive Evaporation and Oxidation of Iron. *Phys. Scr.* **2012**, *85* (5), 055802. https://doi.org/10.1088/0031-8949/85/05/055802.

(43) Bhagya, T. C.; Krishnan, A.; S, A. R.; M, A. S.; Sreelekshmy, B. R.; Jineesh, P.; Shibli, S. M. A. Exploration and Evaluation of Proton Source-Assisted Photocatalyst for Hydrogen Generation. *Photochem. Photobiol. Sci.* **2019**, *18* (7), 1716–1726. https://doi.org/10.1039/C9PP00119K.

(44) Ozaslan, D.; Erken, O.; Gunes, M.; Gumus, C. The Effect of Annealing Temperature on the Physical Properties of Cu2O Thin Film Deposited by SILAR Method. *Phys. B Condens. Matter* **2020**, *580*, 411922. https://doi.org/10.1016/j.physb.2019.411922.

(45) Kim, J. H.; Jeon, K. A.; Kim, G. H.; Lee, S. Y. Electrical, Structural, and Optical Properties of ITO Thin Films Prepared at Room Temperature by Pulsed Laser Deposition. *Appl. Surf. Sci.* **2006**, *252* (13), 4834–4837. https://doi.org/10.1016/j.apsusc.2005.07.134.

(46) Sun, T.; Yu, Y.; Zacher, B. J.; Mirkin, M. V. Scanning Electrochemical Microscopy of Individual Catalytic Nanoparticles. *Angew. Chem. Int. Ed.* **2014**, *53* (51), 14120–14123. https://doi.org/10.1002/anie.201408408.
(47) Thangamuthu, M.; Santschi, C.; Martin, O. J. F. Reliable Langmuir Blodgett Colloidal Masks for Large Area Nanostructure Realization. *Thin Solid Films* **2020**, *709*, 138195. https://doi.org/10.1016/j.tsf.2020.138195.
(48) Harris-Lee, T. R.; Marken, F.; Bentley, C. L.; Zhang, J.; Johnson, A. L. A Chemist's Guide to Photoelectrode Development for Water Splitting – the Importance of Molecular Precursor Design. *EES Catal.* **2023**, *1* (6), 832–873. https://doi.org/10.1039/D3EY00176H.
(49) Huang, X.; Li, Y.; Gao, X.; Xue, Q.; Zhang, R.; Gao, Y.; Han, Z.; Shao, M. The Effect of the Photochemical Environment on Photoanodes for Photoelectrochemical Water Splitting. *Dalton Trans.* **2020**, *49* (35), 12338–12344. https://doi.org/10.1039/D0DT01566K.
(50) Zhang, K.; Ravishankar, S.; Ma, M.; Veerappan, G.; Bisquert, J.; Fabregat-Santiago, F.; Park, J. H. Overcoming Charge Collection Limitation at Solid/Liquid Interface by a Controllable Crystal Deficient Overlayer. *Adv. Energy Mater.* **2017**, *7* (3), 1600923. https://doi.org/10.1002/aenm.201600923.
(51) Xu, Z.; Fan, Z.; Shi, Z.; Li, M.; Feng, J.; Pei, L.; Zhou, C.; Zhou, J.; Yang, L.; Li, W.; Xu, G.; Yan, S.; Zou,
Z. Interface Manipulation to Improve Plasmon-Coupled Photoelectrochemical Water Splitting on α-Fe2O3 Photoanodes. *ChemSusChem* **2018**, *11* (1), 237–244. https://doi.org/10.1002/cssc.201701679.
(52) Singh, A. P.; Kodan, N.; Mehta, B. R.; Held, A.; Mayrhofer, L.; Moseler, M. Band Edge Engineering in BiVO4/TiO2 Heterostructure: Enhanced Photoelectrochemical Performance through Improved Charge Transfer. ACS Catal. 2016, 6 (8), 5311–5318. https://doi.org/10.1021/acscatal.6b00956.
(53) Gao, R.-T.; Zhang, J.; Nakajima, T.; He, J.; Liu, X.; Zhang, X.; Wang, L.; Wu, L. Single-Atomic-Site Platinum Steers Photogenerated Charge Carrier Lifetime of Hematite Nanoflakes for Photoelectrochemical Water Splitting. Nat. Commun. 2023, 14 (1), 2640. https://doi.org/10.1038/s41467-023-38343-6.
(54) Gardner, R. F. G.; Sweett, F.; Tanner, D. W. The Electrical Properties of Alpha Ferric Oxide—II.: Ferric Oxide of High Purity. J. Phys. Chem. Solids 1963, 24 (10), 1183–1196. https://doi.org/10.1016/0022-3697(63)90235-X.
(55) Kennedy, J. H.; Frese, K. W. Photooxidation of Water at α - Fe2 O 3 Electrodes. J. Electrochem. Soc. 1978, 125 (5), 709. https://doi.org/10.1149/1.2131532.
(56) Kiani, F.; Sterl, F.; Tsoulos, T. V.; Weber, K.; Giessen, H.; Tagliabue, G. Ultra-Broadband and Omnidirectional Perfect Absorber Based on Copper Nanowire/Carbon Nanotube Hierarchical Structure. ACS Photonics 2020, 7 (2), 366–374. https://doi.org/10.1021/acsphotonics.9b01658.
(57) Kim, S. J.; Thomann, I.; Park, J.; Kang, J.-H.; Vasudev, A. P.; Brongersma, M. L. Light Trapping for Solar Fuel Generation with Mie Resonances. Nano Lett. 2014, 14 (3), 1446–1452. https://doi.org/10.1021/nl404575e.